\documentclass[envcountsame,envcountsect,orivec]{llncs}

\usepackage[english]{babel}
\usepackage[utf8]{inputenc}
\usepackage{amsmath, amssymb, stmaryrd}

\usepackage{graphicx}

\usepackage{enumitem}
\usepackage{mathpartir}

\usepackage{bussproofs}

\usepackage{xparse} 

\usepackage[usenames,dvipsnames]{xcolor}
\usepackage[colorlinks,
            linkcolor={red!50!black},
            citecolor={blue!50!black},
            urlcolor={blue!80!black}]{hyperref}

\usepackage{tikz-cd}

\usepackage{etoolbox} 

\DeclareSymbolFont{sfoperators}{OT1}{cmss}{m}{n}
\DeclareSymbolFontAlphabet{\mathsf}{sfoperators}

\makeatletter
\def\operator@font{\mathgroup\symsfoperators}
\makeatother

\newcommand{\lambdanext}{\ensuremath{\mathsf{g}\lambda}}
\newcommand{\logiclambdanext}{\ensuremath{L\mathsf{g}\lambda}}
\newcommand{\gdtt}{\ensuremath{\mathsf{gDTT}}}

\newcommand{\typerule}[1]{\textsc{#1}}

\newcommand{\bnfeq}{\mathrel{::=}}
\newcommand{\defeq}{\triangleq}

\newcommand{\judgeq}{\equiv}
\newcommand{\judgeqty}{\simeq}

\renewcommand{\gets}{\shortleftarrow}
\newcommand{\subst}[2]{[#1/#2]}
\newcommand{\esubst}[2]{[#2/#1]} 
\newcommand{\emptyctx}{\mathop{\cdot}}

\newcommand{\NAT}{\mathbb{N}}
\DeclareDocumentCommand{\laterbare}{o}
{\IfNoValueTF{#1}
  {\mathord{\triangleright}}
  {\mathord{\triangleright}^{#1}}}

\DeclareDocumentCommand{\later}{ m o m }{
  \IfNoValueTF{#2}
  {\mathord{\overset{#1}{\triangleright}}#3}
  {\mathord{\overset{#1}{\triangleright}}#2 . #3}}
\newcommand{\vrt}[1]{\left[\begin{array}{l} #1 \end{array}\right]}
\newcommand{\hrt}[1]{\left[ #1 \right]}

\newcommand{\latercode}[1]{\code{\triangleright}^{#1}}
\newcommand{\U}[1]{\mathcal{U}_{#1}}

\newcommand{\El}{\operatorname{El}}

\newcommand{\Unit}{\mathbf{1}}
\newcommand{\Nat}{\ensuremath{\mathbf{N}}}
\newcommand{\Bool}{\ensuremath{\mathbf{B}}}
\newcommand{\constant}{\square}
\newcommand{\idty}[3]{\ensuremath{{\operatorname{Id}_{#1}\hspace{-3pt}\left(#2,#3\right)}}}
\newcommand{\idtyvert}[3]{\ensuremath{{\operatorname{Id}_{#1}\hspace{-3pt}\left(\begin{array}{l}#2,\\#3\end{array}\right)}}}

\newcommand{\depprod}[3]{\ensuremath{{\Pi{\left(#1 : #2\right)}.#3}}}
\newcommand{\depsum}[3]{\ensuremath{{\Sigma\left({#1 : #2}\right)#3}}}
\newcommand{\alwaystype}[2]{\ensuremath{\forall{#1}\ifstrempty{#2}{}{.#2}}}

\newcommand{\alwaysterm}[2]{\ensuremath{\Lambda{#1}\ifstrempty{#2}{}{.#2}}}
\newcommand{\alwaysapp}[2]{\ensuremath{#1\!\left[#2\right]}} 
\newcommand{\alwayscode}{\ensuremath{\operatorname{\code{\forall}}}}

\newcommand{\code}[1]{\widehat{#1}}
\newcommand{\codeop}[1]{\mathop{\widehat{#1}}}
\newcommand{\unit}{\operatorname{\langle\rangle}}

\newcommand{\bif}[3]{\ensuremath{\operatorname{if} #1 \operatorname{then} #2
    \operatorname{else} #3}}

\newcommand{\purebare}{\operatorname{next}}
\DeclareDocumentCommand{\pure}{ m o m }{
  \IfNoValueTF{#2}
  {\operatorname{next}^{#1}#3}
  {\operatorname{next}^{#1} #2 . #3}}
\newcommand{\nextold}[2][]{%
  \ifstrempty{#1}{%
    \operatorname{next} #2}{%
    \operatorname{next} #1 . #2}}
\DeclareDocumentCommand{\app}{o}
{\IfNoValueTF{#1}
  {\ensuremath{\mathbin{\circledast}}}
  {\ensuremath{\mathbin{\ensuremath{\text{\textcircled{\scalebox{0.7}{\ensuremath{#1}}}}}}}}}
\newcommand{\prev}[2]{\operatorname{prev} #1 . #2}
\newcommand{\prevbare}{\operatorname{prev}}

\newcommand{\zero}{0}
\renewcommand{\succ}{\operatorname{succ}}

\newcommand{\pair}[2]{\left\langle #1, #2 \right\rangle}
\DeclareDocumentCommand{\fixcombinator}{o}
{\IfNoValueTF{#1}
  {\operatorname{fix}}
  {\operatorname{fix}^{#1}}
}
\newcommand{\fix}[3]{\fixcombinator[#1] #2 . #3}
\newcommand{\inl}{\operatorname{inl}}
\newcommand{\inr}{\operatorname{inr}}
\newcommand{\caseof}[1]{\ensuremath{\operatorname{case}#1\operatorname{of}}}
\newcommand{\caseinl}[2]{\ensuremath{\inl #1 \Rightarrow #2}}
\newcommand{\caseinr}[2]{\ensuremath{\inr #1 \Rightarrow #2}}
\newcommand{\tm}[1]{\ensuremath{\operatorname{\mathsf{#1}}}}
\newcommand{\refl}[2]{\ensuremath{{\operatorname{r}_{#1}#2}}}

\newcommand{\complus}{\ensuremath{\operatorname{com}^{+}}}
\DeclareDocumentCommand{\comif}{o}
{\IfNoValueTF{#1}
  {\ensuremath{\operatorname{com}^{\operatorname{if}}}}
  {\ensuremath{\operatorname{com}_{#1}^{\operatorname{if}}}}}

\newcommand{\cozero}{\ensuremath{\overline{0}}}
\newcommand{\cosucc}{\ensuremath{\operatorname{\overline{succ}}}}

\DeclareDocumentCommand{\force}{o}
{
  \IfNoValueTF{#1}
  {\ensuremath{\operatorname{force}}}
  {\ensuremath{\operatorname{force}\left(#1\right)}}
}

\newcommand{\comnat}{\ensuremath{\operatorname{com}^{\conat}}}
\newcommand{\covectail}{\ensuremath{\operatorname{tl}}}
\DeclareDocumentCommand{\covecmap}{o}
{\IfNoValueTF{#1}%
  {\ensuremath{\ensuremath{\operatorname{map}}}}%
  {\ensuremath{\ensuremath{\operatorname{map}^{#1}}}}
}

\newcommand{\isclock}[2]{\ensuremath{\vdash_{#1} #2}}
\newcommand{\wfctx}[2]{\ensuremath{#2 \vdash_{#1}}}
\newcommand{\wftype}[3]{\ensuremath{#2 \vdash_{#1} #3 \, \operatorname{type}}}
\newcommand{\hastype}[4]{\ensuremath{#2 \vdash_{#1} #3 : #4}}
\newcommand{\typeeqrel}{\ensuremath{\equiv}}
\newcommand{\typeeq}[4]{\ensuremath{#2 \vdash_{#1} #3 \termeqrel #4}}
\newcommand{\termeqrel}{\ensuremath{\equiv}}
\newcommand{\termeq}[5]{\ensuremath{#2 \vdash_{#1} #3 \termeqrel #4 : #5}}
\newcommand{\typeiso}{\cong}

\newcommand{\ctxmorph}[4]{\ensuremath{{#2 :_{#1} #3 \to #4}}}
\newcommand{\dsubst}[5]{\ensuremath{\vdash_{#1} #3 : #4 \overset{#2}{\rightarrowtriangle} #5}}
\newcommand{\dstocm}[3]{\operatorname{adv}_{#1}^{#2}(#3)}

\newcommand{\Lf}{\ensuremath{\mathfrak{L}}}

\DeclareDocumentCommand{\gstream}{ o m }
{\IfNoValueTF{#1}
  {\ensuremath{\operatorname{Str}^g_{#2}}}
  {\ensuremath{\operatorname{Str}^{#1}_{#2}}}}
\DeclareDocumentCommand{\gtree}{ o m }
{\IfNoValueTF{#1}
  {\ensuremath{\operatorname{Tree}^g_{#2}}}
  {\ensuremath{\operatorname{Tree}^{#1}_{#2}}}}
\newcommand{\stream}[1]{\ensuremath{\operatorname{Str}_{#1}}}

\DeclareDocumentCommand{\lift}{ o m }
{\IfNoValueTF{#1}
  {\ensuremath{#2^g}}
  {\ensuremath{#2^{#1}}}
}

\DeclareDocumentCommand{\hdg}{o}
{\IfNoValueTF{#1}%
  {\ensuremath{\operatorname{hd}^g}}%
  {\ensuremath{\operatorname{hd}^{#1}}}
}
\DeclareDocumentCommand{\tlg}{o}
{\IfNoValueTF{#1}%
  {\ensuremath{\operatorname{tl}^g}}%
  {\ensuremath{\operatorname{tl}^{#1}}}
}

\newcommand{\hd}[1][]{\ensuremath{\operatorname{hd}{#1}}}
\newcommand{\tl}[1][]{\ensuremath{\operatorname{tl}{#1}}}

\DeclareDocumentCommand{\consg}{o}
{\IfNoValueTF{#1}%
  {\ensuremath{\operatorname{cons}^g}}%
  {\ensuremath{\operatorname{cons}^{#1}}}
}

\newcommand{\cons}{\operatorname{cons}}

\DeclareDocumentCommand{\everyother}{o}
{\IfNoValueTF{#1}
  {\ensuremath{\operatorname{eo}}}
  {\ensuremath{\operatorname{eo}^{#1}}}}

\DeclareDocumentCommand{\plusg}{o}
{\IfNoValueTF{#1}
  {\ensuremath{\operatorname{plus}^g}}
  {\ensuremath{\operatorname{plus}^{#1}}}
}

\newcommand{\zipWith}[1]{\ensuremath{\operatorname{zipWith^{#1}}}}

\newcommand{\map}{\ensuremath{\operatorname{map}}}

\newcommand{\pairseta}{\ensuremath{\operatorname{p\eta}}}

\DeclareDocumentCommand{\gconat}{o}
{\IfNoValueTF{#1}%
  {\ensuremath{\operatorname{Co\NAT}}}%
  {\ensuremath{\operatorname{Co\NAT}^{#1}}}
}
\DeclareDocumentCommand{\gcovec}{o m}
{\IfNoValueTF{#1}%
  {\ensuremath{\ensuremath{\operatorname{CoVec}_{#2}}}}%
  {\ensuremath{\ensuremath{\operatorname{CoVec}^{{#1}}_{#2}}}}
}

\newcommand{\conat}{\operatorname{Co\NAT}}
\DeclareDocumentCommand{\covec}{o m}
{\IfNoValueTF{#1}%
  {\ensuremath{\ensuremath{\operatorname{CoVec}_{#2}}}}%
  {\ensuremath{\ensuremath{\operatorname{CoVec}^{#1}_{#2}}}}
}

\DeclareDocumentCommand{\model}{o}
{\IfNoValueTF{#1}
  {\ensuremath{\mathfrak{M}}}
  {\ensuremath{\mathfrak{M}\left(#1\right)}}
}

\newcommand{\clocks}{\ensuremath{\text{CV}}}
\newcommand{\clockconst}{\ensuremath{\kappa_0}}

\newcommand{\comp}{\circ}

\newcommand{\trees}{\ensuremath{\mathcal{S}}}
\newcommand{\latermod}{\ensuremath{\operatorname{\blacktriangleright}}}

\newenvironment{smalldiagram}{\begin{tikzcd}[row sep=1cm,column sep=0.5cm]}{\end{tikzcd}}

\newcommand{\co}{\colon}

\begin{document}
\frontmatter          
\pagestyle{headings}  
\mainmatter              
\title{Guarded Dependent Type Theory with Coinductive Types}
\titlerunning{$\gdtt$}  
%
\author{Ale\v{s} Bizjak\inst{1}\and%
        Hans Bugge Grathwohl\inst{1}\and%
        Ranald Clouston\inst{1}\and \\%
        Rasmus E. M{\o}gelberg\inst{2}\and%
        Lars Birkedal\inst{1}}

\authorrunning{Bizjak \textit{et~al.}} 
\institute{
Aarhus University\\
\email{\{abizjak,hbugge,ranald.clouston,birkedal\}@cs.au.dk}
\and
IT University of Copenhagen\\
\email{mogel@itu.dk}
}

\maketitle              

\begin{abstract}
  We present guarded dependent type theory, $\gdtt$, an extensional dependent type theory with a `later' modality and clock quantifiers for programming and proving with guarded recursive and coinductive types.
  The later modality is used to ensure the productivity of recursive definitions in a modular, type based, way.
  Clock quantifiers are used for controlled elimination of the later modality and for encoding coinductive types using guarded recursive types.
  Key to the development of $\gdtt$ are novel type and term formers involving what we call `delayed substitutions'.
  These generalise the applicative functor rules for the later modality considered in earlier work, and are crucial for programming and proving with dependent types.
  We show soundness of the type theory with respect to a denotational model.

  \emph{This is the technical report version of a paper to appear in the proceedings of FoSSaCS 2016.}
%
\end{abstract}

\section{Introduction}
\label{sec:introduction}

Dependent type theory is useful both for programming, and for proving properties of elements of types.
Modern implementations of dependent type theories such as Coq~\cite{Coq:manual}, Nuprl~\cite{Constable-et-al:nuprl}, Agda~\cite{Norell:thesis}, and Idris~\cite{Brady:idris}, have been
used successfully in many projects.
However, they offer limited support for programming and proving with \emph{coinductive} types.

One of the key challenges is to ensure that functions on
coinductive types are well-defined; that is, productive with unique solutions. Syntactic
guarded recursion~\cite{Coquand:Infinite}, as used for example in Coq~\cite{Gimenez:Codifying},
ensures productivity by
requiring that recursive calls be nested
directly under a constructor, but it is well known that such syntactic checks exclude many 
valid definitions, particularly in the presence of higher-order functions.

To address this challenge, a \emph{type-based} approach to guarded recursion, more flexible than
syntactic checks, was first suggested by Nakano~\cite{Nakano:Modality}.
A new modality, written $\laterbare$ and called `later'~\cite{Appel:Very},
allows us to distinguish between data we have access to now, and data which we will get later.
This modality must be used to guard self-reference in type definitions, so for example \emph{guarded streams} of natural numbers are described by the guarded recursive equation
\[
  \gstream{\NAT}  \judgeqty \NAT\times \laterbare\gstream{\NAT}
\]                              
asserting that stream heads are available now, but tails only later.

Types defined via guarded recursion with $\laterbare$ are not standard coinductive types,
as their denotation is defined via models based on the \emph{topos of
trees}~\cite{Birkedal-et-al:topos-of-trees}. More pragmatically, the bare addition of $\laterbare$
disallows productive but \emph{acausal}~\cite{Krishnaswami:Ultrametric} functions such as the
`every other' function that returns every second element of a stream. Atkey and McBride proposed
\emph{clock quantifiers}~\cite{Atkey:Productive} for such functions; these have been extended to
dependent
types~\cite{Mogelberg:tt-productive-coprogramming,Bizjak-Mogelberg:clock-synchronisation},
and M{\o}gelberg~\cite[Thm. $2$]{Mogelberg:tt-productive-coprogramming} has shown
that they allow the definition of types whose denotation is precisely that of standard coinductive types
interpreted in set-based semantics. As such, they allow us to program with real
coinductive types, while retaining productivity guarantees.

In this paper we introduce the extensional guarded dependent type theory $\gdtt$,
which provides a framework where guarded recursion can be used not just for programming
with coinductive types but also for coinductive reasoning.

As types depend on terms, one of the key challenges in designing $\gdtt$ is coping with
elements that are only available later, i.e., elements of types of the form $\laterbare{A}$.
We do this by generalising the applicative functor structure of $\laterbare$ to the dependent 
setting. Recall the rules for applicative functors~\cite{McBride:Applicative}:
\begin{equation}
    \label{eq:basic-app-typing-old}
    \inferrule*{
      \hastype{}{\Gamma}{t}{A}
    }{
      \hastype{}{\Gamma}{\nextold{t}}{\laterbare A}
    }
    \quad
    \inferrule*{
      \hastype{}{\Gamma}{f}{\laterbare(A\to B)} \\
      \hastype{}{\Gamma}{t}{\laterbare A}
    }{
      \hastype{}{\Gamma}{f\app t}{\laterbare B}
    }
\end{equation}
The first rule allows us to make later use of data that we have now.
The second allows, for example, functions to be applied recursively to the tails of streams.

Suppose now that $f$ has type $\laterbare(\Pi x:A.B)$, and $t$ has type $\laterbare A$.
What should the type of $f\app t$ be? 
Intuitively, $t$ will eventually reduce to some value $\nextold u$,
and so the resulting type should be $\later{}{(B\subst{u}{x})}$, but if
$t$ is an open term we may not be able to perform this reduction.
This problem occurs in coinductive reasoning: if, e.g., $A$ is $\gstream{\NAT}$, and $B$ a property of streams,
in our applications $f$ will be a (guarded) coinduction assumption that we will want to apply
to the tail of a stream, which has type $\laterbare\gstream{\NAT}$.

We hence must introduce a new
notion, of \emph{delayed substitution}, similar to let-binding, allowing us to give $f\app t$ 
the type 
\[
 \later{}[\hrt{x \gets t}]B
\]
binding $x$ in $B$. Definitional equality rules then allow us to simplify
this type when $t$ has form $\nextold{u}$, i.e.,
$\later{}[\hrt{x \gets \nextold{u}}]B \typeeqrel \later{}{(B[u/x])}$.
This construction generalises to bind a list of variables.
Delayed substitution is essential to many examples, as shown in
Sec.~\ref{sec:examples}, and surprisingly the applicative functor term-former $\app$, so central to
the standard presentation of applicative functors, turns out to be \emph{definable} via delayed
substitutions, as shown in Sec.~\ref{sec:gdtt}.

\paragraph{Contributions.}

The contributions of this paper are:
\begin{itemize}
\item We introduce the extensional guarded dependent type theory $\gdtt$, and show that it
gives a framework for programming and proving with guarded recursive and coinductive types.
The key novel feature is the generalisation of the `later' type-former and `next' term-former via
\emph{delayed substitutions};
\item We prove the soundness of $\gdtt$ via a model similar to that used in earlier work on guarded
recursive types and clock
quantifiers~\cite{Mogelberg:tt-productive-coprogramming,Bizjak-Mogelberg:clock-synchronisation}.
\end{itemize}
We focus on the design and soundness of the type theory and restrict attention to an extensional type theory.
We postpone a treatment of an intensional version of the theory to future work (see
Secs.~\ref{sec:related} and~\ref{sec:conclusion}).

In addition to the examples included in this paper, we are pleased to note that a preliminary version of
$\gdtt$ has already proved crucial for formalizing a logical relations adequacy proof of a semantics for PCF using guarded recursive types by
Paviotti~et.~al.~\cite{Paviotti-et-al:PCF}.

\section{Guarded Dependent Type Theory}
\label{sec:gdtt}

$\gdtt$ is a type theory with base types unit $\Unit$, booleans $\Bool$, and natural numbers $\Nat$,
along with $\Pi$-types, $\Sigma$-types, identity types, and universes.
For space reasons we omit all definitions that are standard to such a type theory; see e.g. Jacobs~\cite{Jacobs:cat-logic-type-theory}.
Our universes are \`a la Tarski, so we distinguish between types and terms, and have terms that represent types; they are called \emph{codes} of types and they can be recognised by their circumflex, e.g., $\code\Nat$ is the code of the type $\Nat$.
We have a map $\El$ sending codes of types to their corresponding type.
We follow standard practice and often omit $\El$ in examples, except where it is important to avoid confusion.

\begin{figure}[b]
  \begin{minipage}[c]{0.5\linewidth}
    \begin{align*}
    &{\isclock{\Delta}{\kappa}} && \text{valid clock}\\
    &{\wfctx{\Delta}{\Gamma}} &&\text{well-formed context} \\
    &{\wftype{\Delta}{\Gamma}{A}} &&\text{well-formed type} \\
    &{\hastype{\Delta}{\Gamma}{t}{A}} &&\text{typing judgment}
    \end{align*}
  \end{minipage}
  \hspace{-0.15cm}
  \begin{minipage}[c]{0.5\linewidth}
    \begin{align*}
      &{\typeeq{\Delta}{\Gamma}{A}{B}} &&\text{type equality} \\
      &{\termeq{\Delta}{\Gamma}{t}{u}{A}} &&\text{term equality} \\
      &{\dsubst{\Delta}{\kappa}{\xi}{\Gamma}{\Gamma'}} &&\text{delayed substitution}
    \end{align*}
  \end{minipage}
  \caption{Judgements in \gdtt.}
  \label{fig:judgments}
\end{figure}

We fix a countable set of \emph{clock variables} $\clocks = \{\kappa_1, \kappa_2, \cdots\}$ and a single \emph{clock constant} $\clockconst$,
which will be necessary to define, for example, the function $\hd$ in Sec.~\ref{sec:example-progr-coinductive}.
A \emph{clock} is either a clock variable or the clock constant;
they are intuitively temporal dimensions on which types may depend.
A \emph{clock context} $\Delta, \Delta', \cdots$ is a finite \emph{set} of \emph{clock variables}.
We use the judgement $\isclock{\Delta}{\kappa}$ to express that either $\kappa$ is a clock variable in the set $\Delta$ or $\kappa$ is the clock constant $\clockconst$.
All judgements, summarised in  Fig.~\ref{fig:judgments}, are parametrised by clock contexts.
Codes of types inhabit \emph{universes} $\U{\Delta}$ parametrised by clock contexts similarly.
The universe $\U{\Delta}$ is only well-formed in clock contexts $\Delta'$ where $\Delta \subseteq \Delta'$.
Intuitively, $\U{\Delta}$ contains codes of types that can vary only along dimensions in $\Delta$.
We have \emph{universe inclusions} from $\U{\Delta}$ to $\U{\Delta'}$ whenever $\Delta \subseteq \Delta'$;
in the examples we will not write these explicitly.
Note that we do not have $\code{\U{\Delta}} : \U{\Delta'}$, i.e., these universes do not form a hierarchy.
We could additionally have an orthogonal hierarchy of universes, i.e. for each clock context $\Delta$ a hierarchy of universes $\U{\Delta}^1 : \U{\Delta}^2 : \cdots$.

All judgements are closed under clock weakening and clock substitution.
The former means that if, e.g., $\hastype{\Delta}{\Gamma}{t}{A}$ is derivable then, for any clock variable $\kappa \not\in\Delta$, the judgement $\hastype{\Delta,\kappa}{\Gamma}{t}{A}$ is also derivable.
The latter means that if, e.g., $\hastype{\Delta,\kappa}{\Gamma}{t}{A}$ is derivable and $\isclock{\Delta}{\kappa'}$ then the judgement $\hastype{\Delta}{\Gamma[\kappa'/\kappa]}{t[\kappa'/\kappa]}{A[\kappa'/\kappa]}$ is also derivable, where clock substitution
$[\kappa'/\kappa]$ is defined as obvious.

The rules for guarded recursion can be found in Figs.~\ref{fig:later-next}
and~\ref{fig:type-term-equalities}; rules for coinductive types are postponed until
Sec.~\ref{sec:coinductive-types}.
Recall the `later' type former $\laterbare$, which expresses that something will be available at a later time. In $\gdtt$ we have $\later{\kappa}{}$ for each clock $\kappa$, so we can delay a type along
 different dimensions.
As discussed in the introduction, we generalise the applicative functor structure of each
$\later{\kappa}{}$ via \emph{delayed substitutions}, which allow a substitution to be delayed until its
substituent is available.
We showed in the introduction how a type with a single delayed substitution $\later{\kappa}[\hrt{x \gets t}]{A}$ should work.
However if we have a term $f$ with more than one argument, for example of type $\later{\kappa}{\left(\depprod{x}{A}{\depprod{y}{B}C}\right)}$, and wish to type an application $f \app[\kappa] t \app[\kappa] u$ (where $\app[\kappa]$ is the applicative functor operation $\app$ for clock
$\kappa$) we may have neither $t$ nor $u$ available now, and so we need sequences of delayed substitutions to define the type $\later{\kappa}[\hrt{x \gets t, y \gets u}]{C}$.
Our concrete examples of Sec.~\ref{sec:examples} will show that this issue arises in practice.
We therefore define sequences of delayed substitutions $\xi$.
The new raw types, terms, and delayed substitutions of $\gdtt$ are given by the grammar
\begin{align*}
  \begin{split}
    A, B &\bnfeq \cdots | ~ \later{\kappa}[\xi]{A}
  \end{split}
  \qquad
  \begin{split}
    t, u &\bnfeq \cdots | ~ \pure{\kappa}[\xi]{t} ~|~ \latercode{\kappa} t
  \end{split}
  \qquad
  \begin{split}
    \xi &\bnfeq \emptyctx ~|~ \xi\hrt{x \gets t}.
  \end{split}
\end{align*}
Note that we just write $\later{\kappa}{A}$ where its delayed substitution is the empty
$\emptyctx$, and that $\later{\kappa}[\xi]{A}$ binds the variables substituted for by $\xi$ in
$A$, and similarly for $\purebare$.

The three rules $\typerule{DS-Emp}$, $\typerule{DS-Cons}$, and
$\typerule{Tf-$\later{}{}$}$ are used to construct the type $\later{\kappa}[\xi]{A}$.
These rules formulate how to
generalise these types to arbitrarily long delayed substitutions. Once the type
formation rule is established, the introduction rule $\typerule{Ty-Next}$ is
the natural one.

With delayed substitutions we can \emph{define} $\app[\kappa]$ as
\begin{align*}
  f \app[\kappa] t \defeq \pure{\kappa}[\vrt{g \gets f\\ x \gets t}]{g\,x}.
\end{align*}
Using the rules in Fig.~\ref{fig:later-next} we can derive the following typing judgement for $\app[\kappa]$
\begin{mathpar}
  \inferrule*[right={Ty-$\app$}]
  {
    \hastype{\Delta}{\Gamma}{f}{\later{\kappa}[\xi]{\Pi (x : A) . B}} \\
    \hastype{\Delta}{\Gamma}{t}{\later{\kappa}[\xi]{A}}}
  {%
    \hastype{\Delta}{\Gamma}{f \app[\kappa] t}{\later{\kappa}[\xi[x\gets t]]{B}}
  }
\end{mathpar}

When a term has the form $\pure{\kappa}[\xi\hrt{x \gets \pure{\kappa}[\xi]{u}}]{t}$, then we have
enough information to perform the substitution in both the term and its
type. The rule $\typerule{TmEq-Force}$ applies the substitution by
equating the term with the result of an actual substitution,
$\pure{\kappa}[\xi]{t\subst{u}{x}}$. The rule $\typerule{TyEq-Force}$ does the same for its
type. Using $\typerule{TmEq-Force}$ we can derive the basic term equality 
\[
  (\pure{\kappa}[\xi]{f})\app[\kappa] (\pure{\kappa}[\xi]{t}) \judgeq \pure{\kappa}[\xi]{(ft)}.
\]
typical of applicative functors~\cite{McBride:Applicative}.

It will often be the case that a delayed substitution is unnecessary, because
the variable to be substituted for does not occur free in the type/term. This is
what $\typerule{TyEq-$\later{}{}$-Weak}$ and $\typerule{TmEq-Next-Weak}$
express, and with these we can justify the simpler typing rule
\[
  \inferrule*
  {\hastype{\Delta}{\Gamma}{f}{\later{\kappa}[\xi]{(A \to B)}} \\
    \hastype{\Delta}{\Gamma}{t}{\later{\kappa}[\xi]{A}}}
  {\hastype{\Delta}{\Gamma}{f \app[\kappa] t}{\later{\kappa}[\xi]{B}}}
\]
In other words, delayed substitutions on the
type are not necessary when we apply a non-dependent function. 

Further, we have the applicative functor identity law
\[
  (\pure{\kappa}[\xi]{\lambda x . x}) \app[\kappa] t \judgeq t.
\]
This follows from the rule $\typerule{TmEq-Next-Var}$, which allows us to
simplify a term $\pure{\kappa}[\xi\hrt{y \gets t}]{y}$ to $t$.

Sometimes it is necessary to switch the order in the delayed substitution.  Two
substitutions can switch places, as long as they do not depend on each other; this is what
$\typerule{TyEq-$\later{}{}$-Exch}$ and $\typerule{TmEq-Next-Exch}$ express.

Rule \textsc{TmEq-Next-Comm} is not used in the examples of this paper, but it implies the rule 
$\pure{\kappa}[\xi\hrt{x\gets t}]{\pure{\kappa}x} \termeqrel \pure{\kappa}{t}$, which 
is needed in Paviotti's PhD work.  

\begin{figure}[t]
  \paragraph{Universes}
  \begin{mathpar}
    \inferrule*[right={Univ}]
    {
      \Delta' \subseteq \Delta \\ \wfctx{\Delta}{\Gamma}
    }
    {%
      \wftype{\Delta}{\Gamma}{\U{\Delta'}}
    }
    \and
    \inferrule*[right={El}]
    {
      \hastype{\Delta}{\Gamma}{A}{\U{\Delta'}}
    }
    {%
      \wftype{\Delta}{\Gamma}{\El(A)}
    }
  \end{mathpar}
  \paragraph{Delayed substitutions:} 
  \begin{mathpar}
    \inferrule*[right={DS-Emp}]
    {
      \wfctx{\Delta}{\Gamma} \\
      \isclock{\Delta}{\kappa}
    }
    {
      \dsubst{\Delta}{\kappa}{\emptyctx}{\Gamma}{\emptyctx}
    }
    \and
    \inferrule*[right={DS-Cons}]
    {
      \dsubst{\Delta}{\kappa}{\xi}{\Gamma}{\Gamma'} \\
      \hastype{\Delta}{\Gamma}{t}{\later{\kappa}[\xi]{A}}
    }{
      \dsubst{\Delta}{\kappa}{\xi\hrt{x\gets t}}{\Gamma}{\Gamma', x : A}
    }
  \end{mathpar}
  
  \paragraph{Typing rules:}
  \begin{mathpar}
    \inferrule*[right=Tf-$\later{}{}$]
    {
      \wftype{\Delta}{\Gamma,\Gamma'}{A} \\
      \dsubst{\Delta}{\kappa}{\xi}{\Gamma}{\Gamma'}
    }
    {%
      \wftype{\Delta}{\Gamma}{\later{\kappa}[\xi]{A}}
    }
    \and
    \inferrule*[right={Ty-$\latercode{}$}]
    {
      \isclock{\Delta'}{\kappa} \\
      \hastype{\Delta}{\Gamma}{A}{\later{\kappa}{\U{\Delta'}}}
    }
    {%
      \hastype{\Delta}{\Gamma}{\latercode{\kappa} A}{\U{\Delta'}}
    }
    \and
    \inferrule*[right={Ty-Next}]
    {
      \hastype{\Delta}{\Gamma,\Gamma'}{t}{A} \\
      \dsubst{\Delta}{\kappa}{\xi}{\Gamma}{\Gamma'}
    }
    {%
      \hastype{\Delta}{\Gamma}{\pure{\kappa}[\xi]{t}}{\later{\kappa}[\xi]{A}}
    }
    \and
    \inferrule*[right={Ty-Fix}]
    {\isclock{\Delta}{\kappa} \\
      \hastype{\Delta}{\Gamma, x : \later{\kappa}{A}}{t}{A}
    }
    {%
      \hastype{\Delta}{\Gamma}{\fix{\kappa}{x}{t}}{A}
    }
  \end{mathpar}
  \caption{Overview of the new typing rules involving $\triangleright$ and delayed substitutions.}
  \label{fig:later-next}
\end{figure}

\begin{figure}[t]
  \paragraph{Definitional type equalities:}
  \begin{align*}
    \later{\kappa}[\xi\hrt{x\gets t}]A
    &\typeeqrel
      \later{\kappa}[\xi]{A}
    && \text{\sc (TyEq-$\later{}{}$-Weak)} \\
    \later{\kappa}[\xi\hrt{x\gets t,y\gets u}\xi']{A}
    &\typeeqrel
      \later{\kappa}[\xi\hrt{y\gets u,x\gets t}\xi']{A}
    && \text{\sc (TyEq-$\later{}$-Exch)} \\
    \later{\kappa}[\xi\hrt{x\gets\pure{\kappa}[\xi]{t}}]{A}
    & \typeeqrel
      \later{\kappa}[\xi]{A\subst{t}{x}}
    && \text{\sc (TyEq-Force)} \\
    \El(\latercode{\kappa}\left(\pure{\kappa}[\xi]{t}\right))
    & \typeeqrel
      \later{\kappa}[\xi]\El(t)
    && \text{\sc (TyEq-El-$\later{}$)} \\
    \idty{\later{\kappa}[\xi]{A}}{\pure{\kappa}[\xi]{t}}{\pure{\kappa}[\xi]{s}}
    & \typeeqrel
      \later{\kappa}[\xi]{\idty{A}{t}{s}}
    && \text{\sc (TyEq-$\later{}{}$)}
  \end{align*}

  \paragraph{Definitional term equalities:}

  \begin{align*}
    \pure{\kappa}[\xi\hrt{x \gets t}]{u}
    & \termeqrel
      \pure{\kappa}[\xi]{u}
    && \text{\sc (TmEq-Next-Weak)} \\
    \pure{\kappa}[\xi\hrt{x\gets t}]{x}
    & \termeqrel
      t
    && \text{\sc (TmEq-Next-Var)} \\
    \pure{\kappa}[\xi\hrt{x \gets t, y \gets u}\xi']{v}
    & \termeqrel
      \pure{\kappa}[\xi\hrt{y \gets u, x \gets t}\xi']{v}
    && \text{\sc (TmEq-Next-Exch)} \\
    \pure{\kappa}[\xi]{\pure{\kappa}[\xi']{u}}
    & \termeqrel
    \pure{\kappa}[\xi']{\pure{\kappa}[\xi]{u}}
    && \text{\sc (TmEq-Next-Comm)} \\
    \pure{\kappa}[\xi\hrt{x \gets \pure{\kappa}[\xi]{t}}]{u}
    & \termeqrel
      \pure{\kappa}[\xi]{u\subst{t}{x}}
    && \text{\sc (TmEq-Force)} \\
    \fix{\kappa}{x}{t}
    & \termeqrel
      t\subst{\pure{\kappa}{\left(\fix{\kappa}{x}{t}\right)}}{x}
    && \text{\sc (TmEq-Fix)}
  \end{align*}
  \caption{New type and term equalities in $\gdtt$.
    Rules \textsc{TyEq-$\later{}{}$-Weak} and \textsc{TmEq-Next-Weak} require that $A$ and $u$ are well-formed in a context without $x$.
    Rules \textsc{TyEq-$\later{}$-Exch} and \textsc{TmEq-Next-Exch} assume that exchanging $x$ and $y$ is allowed, i.e., that the type of $x$ does not depend on $y$ and vice versa.
    Likewise, rule \textsc{TmEq-Next-Comm} assumes that exchanging the codomains of $\xi$ and $\xi'$ is allowed and that none of the variables in the codomains of $\xi$ and $\xi'$ appear in the type of $u$.}
  \label{fig:type-term-equalities}
\end{figure}

\subsection{Fixed points and guarded recursive types}
\label{sec:fixed-points}

In $\gdtt$ we have for each clock $\kappa$ valid in the current clock context a fixed-point combinator $\fixcombinator[\kappa]$.
This differs from a traditional fixed-point combinator in that the type of the recursion variable is not the same as the result type;
instead its type is \emph{guarded} with $\later{\kappa}{}$.
When we define a term using the fixed-point, we say that it is defined by \emph{guarded recursion}.
When the term is intuitively a proof, we say we are proving by
\emph{L\"ob induction}~\cite{Appel:Very}.

\emph{Guarded recursive types} are defined as fixed-points of suitably guarded functions on universes.
This is the approach of Birkedal and M{\o}gel\-berg~\cite{Birkedal:intensional}, but the generality of the rules of $\gdtt$ allows us to define more interesting dependent guarded recursive types, for example the predicates of Sec.~\ref{sec:examples}.

We first illustrate the technique by defining the (non-dependent) type of guarded streams.
Recall from the introduction that we want the type of guarded streams, for clock $\kappa$, to satisfy the equation $\gstream[\kappa]{A} \typeeqrel A \times \later{\kappa}{\gstream[\kappa]{A}}$.

The type $A$ will be equal to $\El(B)$ for some code $B$ in some universe $\U{\Delta}$ where the
clock variable $\kappa$ is not in $\Delta$.
We then define the \emph{code} $S_A^\kappa$ of $\gstream[\kappa]{A}$ in the universe $\U{\Delta, \kappa}$ to be
$S_A^\kappa \defeq \fix{\kappa}{X}{B \codeop\times \latercode{\kappa} X}$,
where $\code\times$ is the code of the (simple) product type.
Via the rules of $\gdtt$ we can show
$\gstream[\kappa]{A} \judgeqty A \times \later{\kappa}\gstream[\kappa]{A}$ as desired.

The head and tail operations, $\hdg[\kappa] : \gstream[\kappa]{A} \to A$ and $\tlg[\kappa] : \gstream[\kappa]{A} \to \later{\kappa}{\gstream[\kappa]{A}}$ are simply the first and the second projections.
Conversely, we construct streams by pairing.
We use the suggestive $\consg[\kappa]$ notation which we define as
\begin{align*}
  \begin{split}
    \consg[\kappa] &: A \to \later{\kappa}\gstream[\kappa]{A} \to \gstream[\kappa]{A}
  \end{split}
  \qquad
  \begin{split}
    \consg[\kappa] &\defeq \lambda \left(a : A\right) \left(as :
      \later{\kappa}\gstream[\kappa]{A}\right) . \pair{a}{as}
  \end{split}
\end{align*}
Defining guarded streams is also done via guarded recursion, for example the stream consisting only of ones is defined as $\tm{ones} \defeq \fix{\kappa}{x}{\consg[\kappa] 1\, x}$.

The rule $\typerule{TyEq-El-$\later{}{}$}$ is essential for defining guarded recursive types
as fixed-points on universes, and it can also be used for defining more advanced guarded
recursive dependent types such as covectors; see Sec.~\ref{sec:examples}.

\subsection{Identity types}
\label{sec:identity-types}

$\gdtt$ has standard extensional identity types $\idty{A}{t}{u}$ (see, e.g., Jacobs~\cite{Jacobs:cat-logic-type-theory}) but with two additional type equivalences necessary for working with guarded dependent types.
We write $\refl{A}t$ for the reflexivity proof $\idty{A}{t}{t}$.
The first type equivalence is the rule \typerule{TyEq-$\later{}{}$}.
This rule, which is validated by the model of Sec.~\ref{sec:soundness}, may be thought of by analogy to type equivalences often considered in homotopy type theory~\cite{hottbook}, such as
\begin{align}
  \label{eq:type-equality-products-example}
  \idty{A \times B}{\pair{s_1}{s_2}}{\pair{t_1}{t_2}}
  \typeeqrel
  \idty{A}{s_1}{t_1} \times \idty{B}{s_2}{t_2}.
\end{align}
There are two important differences.
The first is that \eqref{eq:type-equality-products-example} is (using univalence) a propositional type equality, whereas \typerule{TyEq-$\later{}{}$} specifices a definitional type equality.
This is natural in an extensional type theory.
The second difference is that there are terms going in both directions in \eqref{eq:type-equality-products-example}, whereas we would have a term of type
$\idty{\later{\kappa}[\xi]{A}}{\pure{\kappa}[\xi]{t}}{\pure{\kappa}[\xi]{u}}
\to
\later{\kappa}[\xi]{\idty{A}{t}{u}}$ without the rule \typerule{TyEq-$\later{}{}$}.

The second novel type equality rule, which involves clock quantification, will be presented in Sec.~\ref{sec:coinductive-types}.

\section{Examples}
\label{sec:examples}

In this section we present some example terms typable in $\gdtt$. Our examples will use a term, which
we call $\pairseta$, of type $ \Pi(s, t : A \times B) . \idty{A}{\pi_1 t}{\pi_1 s} \to
\idty{B}{\pi_2 t}{\pi_2 s} \to
\idty{A \times B}{t}{s}$.
This term is definable in any type theory with a strong (dependent) elimination rule for dependent sums.
The second property we will use is that $\gstream[\kappa]{A} \typeeqrel A \times \later{\kappa}\gstream[\kappa]{A}$.
Because $\hdg[\kappa]$ and $\tlg[\kappa]$ are simply first and second projections, $\pairseta$ also has type
$ \Pi \left(xs,ys : \gstream[\kappa]{A}\right).
{\idty{A}{\hdg[\kappa]{xs}}{\hdg[\kappa]{ys}}} \to
{\idty{\later{\kappa}\gstream[\kappa]{A}}{\tlg[\kappa]{xs}}{\tlg[\kappa]{ys}}} 
\to {\idty{\gstream[\kappa]{A}}{xs}{ys}}$.

\paragraph{$\zipWith{\kappa}$ preserves commutativity.}
\label{sec:zipwith-preserves-comm}

In $\gdtt$ we define the $\zipWith{\kappa}$ function which has the type $(A \to
B \to C) \to \gstream[\kappa]{A} \to \gstream[\kappa]{B} \to
\gstream[\kappa]{C}$ by
\begin{align*}
  \zipWith{\kappa} f &\defeq \fix{\kappa}{\phi}{\lambda xs, ys  . 
                    \consg[\kappa] \left(f\, (\hdg[\kappa]{xs})\,(\hdg[\kappa]{ys})\right)
                    \left(\phi\app[\kappa] \tlg[\kappa]{xs} \app[\kappa] \tlg[\kappa]{ys}\right)}.
\end{align*}
We show that commutativity of $f$ implies commutativity of $\zipWith{\kappa} f$,
i.e., that
\begin{align*}
  &\Pi (f : A \to A \to B) . \left(\depprod{x,y}{A}{\idty{B}{f\,x\,y}{f\,y\,x}}\right) \to\\
  &\depprod{xs,ys}{\gstream[\kappa]{A}}
  {\idty{\gstream[\kappa]{B}}{\zipWith{\kappa} f\,xs\,ys}{\zipWith{\kappa} f\,ys\,xs}}
\end{align*}
is inhabited.
The term that inhabits this type is 
\begin{align*}
  \lambda f.\lambda c.\fix{\kappa}{\phi}{\lambda xs, ys .
  \pairseta\,\left(c\,(\hdg[\kappa]{xs})\,(\hdg[\kappa]{ys})\right)
                      \,\left(\phi\app[\kappa] \tlg[\kappa]{xs}\app[\kappa] \tlg[\kappa]{ys}\right)}.
\end{align*}
Here, $\phi$ has type $\later\kappa(\depprod{xs,ys}{\gstream[\kappa]{A}}
  {\idty{\gstream[\kappa]{B}}{\zipWith{\kappa} f\,xs\,ys}{\zipWith{\kappa} f\,ys\,xs}})$
so to type the term above, we crucially need delayed substitutions. 

\paragraph{An example with covectors.}
\label{sec:an-example-with-covectors}

The next example is more sophisticated, as it involves programming and proving with a data type that, unlike streams, is dependently typed.
Indeed the generalised later, carrying a delayed substitution, is necessary to type even elementary programs.
\emph{Covectors} are the potentially infinite version of vectors (lists with length).
To define guarded covectors we first need guarded co-natural numbers.
The definition in $\gdtt$ is $\gconat[\kappa] \defeq \El\left(\fix{\kappa}{X}{(\code{\Unit} \codeop{+} \latercode{\kappa}X)}\right)$;
this type satisfies
$\gconat[\kappa] \typeeqrel \Unit + \later{\kappa}\gconat[\kappa]$.
Using $\gconat[\kappa]$ we can define the type family of covectors
$\gcovec[\kappa]{A} n \defeq \El(\code{\gcovec[\kappa]{A}}\,n)$, where 
\begin{align*}
  \code{\gcovec[\kappa]{A}} \defeq
  &\fix{\kappa}{\left(\phi : \later{\kappa}{(\gconat[\kappa] \to \U{\Delta,\kappa})}\right)}{\lambda (n : \gconat[\kappa]).
  \caseof{n}\\
  &\caseinl{u}{\code{\Unit}}\\
  &\caseinr{m}{A \codeop{\times} \latercode{\kappa}(\phi \app[\kappa] m)}}.
\end{align*}
We will not distinguish between $\gcovec[\kappa]{A}$ and $\code{\gcovec[\kappa]{A}}$.
As an example of covectors, we define $\tm{ones}$ of type $\Pi (n : \gconat[\kappa]) .
\gcovec[\kappa]{\NAT} n$ which produces a covector of any length consisting only of ones:
\begin{align*}
  \tm{ones} \defeq 
  \fix{\kappa}{\phi}{\lambda (n : \gconat[\kappa]) . \caseof{n}}\left\{ \caseinl{u}{\inl{\unit}}; \caseinr{m}{\pair{1}{\phi \app[\kappa] m}}\right\}.
\end{align*}
Although this is one of the simplest covector programs one can imagine, it does not type-check without
the generalised later with delayed substitutions.

The $\map$ function on covectors is defined as
\begin{align*}
  \map ~:~& (A \to B) \to \depprod{n}{\gconat[\kappa]}{\gcovec[\kappa]{A}{n} \to \gcovec[\kappa]{B}{n}}\\
  \map f \defeq\,
  &\fix{\kappa}{\phi}{\lambda (n : \gconat[\kappa]) .
  \caseof{n}\\
  &\caseinl{u}{\lambda (x : 1) . x}\\
  &\caseinr{m}{\lambda\left(p : A \times \later{\kappa}[\hrt{n\gets m}](\gcovec[\kappa]{A} n)\right) .
                 \pair{f\left(\pi_1 p\right)}{\phi \app[\kappa] m \app[\kappa] (\pi_2p)}
               }}.
\end{align*}
It preserves composition: the following type is inhabited
\begin{align*}
  \begin{split}
    &\Pi(f : A \to B) (g : B \to C) (n : \gconat[\kappa]) (xs : \gcovec[\kappa]{A} n) .\\
    &\qquad\idty{\gcovec[\kappa]{C}n}{\map g\, n\,(\map f\,n\,xs)}{\map\, (g \comp f)\,n\,xs}
  \end{split}
\end{align*}
by the term
\begin{align*}
  \lambda &(f : A \to B) (g : B \to C) . \fix{\kappa}{\phi}{\lambda (n : \gconat[\kappa]).
  \caseof{n}\\
  &\caseinl{u}{\lambda (xs : 1) . \refl{1}{xs}}\\
  &\caseinr{m}{
  \lambda (xs : \gcovec[\kappa]{A}(\inr m)) .
   \pairseta \left(\refl{C}{g(f(\pi_1xs))}\right)
    \left(\phi\app[\kappa] m\app[\kappa] \pi_2 xs\right)}}.
\end{align*}

\section{Coinductive types}
\label{sec:coinductive-types}


As discussed in the introduction, guarded recursive types on their own disallow productive but acausal function definitions.
To capture such functions we need to be able to remove $\later{\kappa}{}$.
However such eliminations must be controlled to avoid trivialising $\later{\kappa}{}$.
If we had an unrestricted elimination term $\operatorname{elim} : \later{\kappa}{A} \to A$ every type would be inhabited via $\fixcombinator[\kappa]$, making the type theory inconsistent.

However, we may eliminate $\later{\kappa}{}$ provided that the term does not depend on the clock $\kappa$, i.e., the term is typeable in a context where $\kappa$ does not appear.
Intuitively, such contexts have no temporal properties along the $\kappa$ dimension, so we may progress the computation without violating guardedness.
Fig.~\ref{fig:coinductive_rules} extends the system of Fig.~\ref{fig:later-next} to allow the removal of clocks in such a setting, by introducing \emph{clock quantifiers} $\alwaystype{\kappa}{}$~\cite{Atkey:Productive,Mogelberg:tt-productive-coprogramming,Bizjak-Mogelberg:clock-synchronisation}.
This is a binding construct with associated term constructor $\alwaysterm{\kappa}{}$, which also binds $\kappa$.
The elimination term is \emph{clock application}.
Application of the term $t$ of type $\alwaystype{\kappa}{A}$ to a clock $\kappa$ is written as $\alwaysapp{t}{\kappa}$.
One may think of $\alwaystype{\kappa}{A}$ as analogous to the type $\forall\alpha.A$ in polymorphic lambda calculus; indeed the basic rules are precisely the same, but we have an additional construct $\prev{\kappa}{t}$, called `previous', to allow removal of the later modality $\later{\kappa}{}$.

Typing this new construct $\prev{\kappa}{t}$ is somewhat complicated, as
it requires `advancing' a delayed substitution, which turns it into a context morphism (an actual substitution); see Fig.~\ref{fig:advancing-delayed} for the definition.
The judgement $\ctxmorph{\Delta}{\rho}{\Gamma}{\Gamma'}$ expresses that $\rho$ is a context morphism from context $\wfctx{\Delta}{\Gamma}$ to the context $\wfctx{\Delta}{\Gamma'}$.
We use the notation $\rho\esubst{x}{t}$ for extending the context morphism by mapping the variable $x$ to the term $t$.
We illustrate this with two concrete examples.

First, we can indeed remove later under a clock quantier:
\begin{align*}
  \begin{split}
    \force &: \alwaystype{\kappa}{\later{\kappa}{A}} \to \alwaystype{\kappa}{A}
  \end{split}
  \begin{split}
    \force &\defeq \lambda x.\prev{\kappa}{\alwaysapp{x}{\kappa}}.
  \end{split}
\end{align*}
The type is correct because advancing the empty delayed substitution in $\later{\kappa}$ turns it into the identity substitution $\iota$, and $A\iota \typeeqrel A$.
The $\beta$ and $\eta$ rules ensure that $\force$ is the inverse to the canonical term $\lambda x.\alwaysterm{\kappa}{\pure{\kappa}{\alwaysapp{x}{\kappa}}}$ of type $\alwaystype{\kappa}{A} \to \alwaystype{\kappa}{\later{\kappa}{A}}$.

Second, we may see an example with a non-empty delayed substitution in the term 
${\prev{\kappa}{\pure{\kappa}{\lambda n.\succ n} \app[\kappa] \pure{\kappa}{\zero}}}$
of type $\alwaystype{\kappa}{\NAT}$.
Recall that $\app[\kappa]$ is syntactic sugar and so more precisely the term is
\begin{align}
  \label{eq:previous-example}
  \prev{\kappa}{\pure{\kappa}[\vrt{f\gets \pure{\kappa}{\lambda n.\succ n}\\ x \gets \pure{\kappa}{\zero}}]{f\,x}}.
\end{align}
Advancing the delayed substitution turns it into the substitution mapping the variable $f$ to the term $\alwaysapp{(\prev{\kappa}{\pure{\kappa}{\lambda n.\succ n}})}{\kappa}$ and the variable $x$ to the term $\alwaysapp{(\prev{\kappa}{\pure{\kappa}{\zero}})}{\kappa}$.
Using the $\beta$ rule for $\prevbare$, then the $\beta$ rule for $\forall\kappa$, this simplifies to the substitution mapping $f$ to $\lambda n.\succ n$ and $x$ to $\zero$.
With this we have that the term \eqref{eq:previous-example} is equal to
$\alwaysterm{\kappa}\left((\lambda n.\succ n)\,\zero\right)$ which is in turn equal to $\alwaysterm{\kappa}{1}$.

An important property of the term $\prev{\kappa}{t}$ is that $\kappa$ is \emph{bound} in $t$; hence $\prev{\kappa}{t}$ has type $\alwaystype{\kappa}{A}$ instead of just $A$.
This ensures that substitution of terms in types and terms is well-behaved and we do not need the explicit substitutions used, for example, by Clouston~et~al.~\cite{Clouston:Programming} where the unary type-former $\constant$
was used in place of clocks.
This binding structure ensures, for instance, that the introduction rule \typerule{Ty-$\Lambda$} closed under substitution in $\Gamma$.

The rule \typerule{TmEq-$\forall$-fresh} states that if $t$ has type $\alwaystype{\kappa}{A}$ and the clock $\kappa$ does not appear in the \emph{type} $A$, then it does not matter to which clock $t$ is applied, as the resulting term will be the same.
In the polymorphic lambda calculus, the corresponding rule for universal quantification over types would be a consequence
of relational parametricity.

We further have the construct $\alwayscode$ and the rule \typerule{Ty-$\forall$-code} which witness that the universes are closed under $\forall\kappa$.

To summarise, the new raw types and terms, extending those of Sec.~\ref{sec:gdtt}, are
\begin{align*}
  \begin{split}
    A, B &\bnfeq \cdots | ~ \alwaystype{\kappa}{A}
  \end{split}
  \begin{split}
    t, u &\bnfeq \cdots | ~ \alwaysterm{\kappa}{t} ~|~ \alwaysapp{t}{\kappa} ~|~
    \alwayscode{t} ~|~ \prev{\kappa}{t}
  \end{split}
\end{align*}

Finally, we have the equality rule \typerule{TyEq-$\forall$-Id} analogous to the rule \typerule{TyEq-$\later{}{}$}.
Note that, as in Sec.~\ref{sec:identity-types}, there is a canonical term of type $\idty{\alwaystype{\kappa}{A}}{t}{s} \to \alwaystype{\kappa}{\idty{A}{\alwaysapp{t}{\kappa}}{\alwaysapp{s}{\kappa}}}$ but, without this rule, no term in the reverse direction.

\begin{figure}[t]
  \begin{mathpar}
  \inferrule*[Right=Tf-$\forall$]
  {
    \wfctx{\Delta}{\Gamma} \\
    \wftype{\Delta,\kappa}{\Gamma}{A}
  }
  {
    \wftype{\Delta}{\Gamma}{\alwaystype{\kappa}{A}}
  }
  \and
  \inferrule*[Right=Ty-$\forall$-code]
  {
    \Delta'\subseteq \Delta \\
    \hastype{\Delta}{\Gamma}{t}{\alwaystype{\kappa}{\U{\Delta',\kappa}}}
  }
  {
    \hastype{\Delta}{\Gamma}{\alwayscode t}{\U{\Delta'}}
  }
  \hspace{2cm}
  \and
  \inferrule*[Right=Ty-$\Lambda$]
  {
    \wfctx{\Delta}{\Gamma} \\
    \hastype{\Delta,\kappa}{\Gamma}{t}{A}
  }
  {
    \hastype{\Delta}{\Gamma}{\alwaysterm{\kappa}{t}}{\alwaystype{\kappa}{A}}
  }
  \and
  \inferrule*[Right=Ty-app]
  {
    \isclock{\Delta}{\kappa'} \\
    \hastype{\Delta}{\Gamma}{t}{\alwaystype{\kappa}{A}}\\
  }
  {
    \hastype{\Delta}{\Gamma}{\alwaysapp{t}{\kappa'}}{A[\kappa'/\kappa]}
  }
  \and
  \inferrule*[Right=Ty-$\operatorname{prev}$]
  { 
    \wfctx{\Delta}{\Gamma} \\
    \hastype{\Delta, \kappa}{\Gamma}{t}{\later{\kappa}[\xi]{A}}
  }
  {
    \hastype{\Delta}{\Gamma}{\prev{\kappa}{t}}{\alwaystype{\kappa}{(A(\dstocm{\Delta}{\kappa}{\xi}))}}
  }
  \end{mathpar}
  \caption{Overview of the new typing rules for coinductive types.}
  \label{fig:coinductive_rules}
\end{figure}

\begin{figure}[t]
  \centering
  \begin{mathpar}
  \inferrule*
  {
    \dsubst{\Delta,\kappa}{\kappa}{\emptyctx}{\Gamma}{\emptyctx} \\
    \wfctx{\Delta}{\Gamma}
  }{
    \dstocm{\Delta}{\kappa}{\emptyctx}
    \defeq
    \ctxmorph{\Delta, \kappa}{\iota}{\Gamma}{\Gamma}
  }
  \and
  \inferrule*
  {
    \dsubst{\Delta,\kappa}{\kappa}{\xi[x\gets t]}{\Gamma}{\Gamma', x : A} \\
    \wfctx{\Delta}{\Gamma}
  }{
    \dstocm{\Delta}{\kappa}{\xi[x \gets t]}
    \defeq
    \ctxmorph{\Delta,\kappa}{\dstocm{\Delta}{\kappa}{\xi}\esubst{x}{\alwaysapp{(\prev{\kappa}{t})}{\kappa}}}
             {\Gamma}{\Gamma,\Gamma',x : A}
  }
  \end{mathpar}
  \caption{Advancing a delayed substitution.}
  \label{fig:advancing-delayed}
\end{figure}

\begin{figure}[t]
    \paragraph{Definitional type equalities:}
  \begin{mathpar}
    \inferrule*[Right=TyEq-$\forall$-el]
    {
      \wfctx{\Delta}{\Gamma} \\
      \Delta' \subseteq \Delta \\
      \hastype{\Delta, \kappa}{\Gamma}{t}{\U{\Delta', \kappa}}
    }
    {
      \typeeq{\Delta}{\Gamma}{\El(\alwayscode \alwaysterm{\kappa}{t})}{\alwaystype{\kappa}{\El(t)}}
    }

    \and

    \inferrule*[Right=TyEq-$\forall$-Id]
    {
      \wfctx{\Delta}{\Gamma}\\
      \wftype{\Delta,\kappa}{\Gamma}{A}\\
      \hastype{\Delta}{\Gamma}{t}{\alwaystype{\kappa}{A}}\\
      \hastype{\Delta}{\Gamma}{s}{\alwaystype{\kappa}{A}}
    }
    {
      \typeeq{\Delta}{\Gamma}
      {\alwaystype{\kappa}{\idty{A}{\alwaysapp{t}{\kappa}}{\alwaysapp{s}{\kappa}}}}
      {\idty{\alwaystype{\kappa}{A}}{t}{s}}
    }
  \end{mathpar}

  \paragraph{Definitional term equalities:}
  \begin{mathpar}
    \inferrule*[Right={TmEq-$\forall$-$\beta$}]
    {
      \wfctx{\Delta}{\Gamma} \\
      \isclock{\Delta}{\kappa'} \\
      \hastype{\Delta, \kappa}{\Gamma}{t}{A}
    }
    {    \termeq{\Delta}{\Gamma}{\alwaysapp{(\alwaysterm{\kappa}{t})}{\kappa'}}{t[\kappa'/\kappa]}{A[\kappa'/\kappa]}
    }

    \and

    \inferrule*[Right=TmEq-$\forall$-$\eta$]
    {
      \kappa \not\in \Delta \\
      \hastype{\Delta}{\Gamma}{t}{\alwaystype{\kappa}{A}}
    }
    {
      \termeq{\Delta}{\Gamma}{\alwaysterm{\kappa}{\alwaysapp{t}{\kappa}}}{t}{\alwaystype{\kappa}{A}}
    }

    \and

    \inferrule*[Right=TmEq-$\forall$-fresh]
    {
      \kappa \not\in \Delta \\
      \wftype{\Delta}{\Gamma}{A} \\
      \hastype{\Delta}{\Gamma}{t}{\alwaystype{\kappa}{A}}\\
      \isclock{\Delta}{\kappa'}\\
      \isclock{\Delta}{\kappa''}
    }
    {
      \termeq{\Delta}{\Gamma}{\alwaysapp{t}{\kappa'}}{\alwaysapp{t}{\kappa''}}{A}
    }

    \and

    \inferrule*[Right=TmEq-$\operatorname{prev}$-$\beta$]
    { 
      \wfctx{\Delta}{\Gamma} \\
      \dsubst{\Delta, \kappa}{\kappa}{\xi}{\Gamma}{\Gamma'} \\
      \hastype{\Delta,\kappa}{\Gamma, \Gamma'}{t}{A}
    }
    {
      \termeq{\Delta}{\Gamma}
      {\prev{\kappa}{\pure{\kappa}[\xi]{t}}}
      {\alwaysterm{\kappa}{t(\dstocm{\Delta}{\kappa}{\xi})}}
      {\alwaystype{\kappa}{(A(\dstocm{\Delta}{\kappa}{\xi}))}}
    }
    
    \and

    \inferrule*[Right=TmEq-$\operatorname{prev}$-$\eta$]
    {
      \wfctx{\Delta}{\Gamma} \\
      \hastype{\Delta, \kappa}{\Gamma}{t}{\later{\kappa}{A}}
    }
    {
      \termeq{\Delta, \kappa}
      {\Gamma}
      {\pure{\kappa}{\left(\alwaysapp{(\prev{\kappa}{t})}{\kappa}\right)}}
      {t}
      {\later{\kappa}{A}}
    }
  \end{mathpar}
  \caption{Type and term equalities involving clock quantification.}
  \label{fig:type-term-eqs-clocks}
\end{figure}

\subsection{Derivable type isomorphisms}
\label{sec:deriv-type-isom}

The encoding of coinductive types using guarded recursive types crucially uses a family of type isomorphisms commuting $\forall\kappa$ over other type formers~\cite{Atkey:Productive,Mogelberg:tt-productive-coprogramming}.
By a type isomorphism $A \typeiso B$ we mean two well-typed terms $f$ and $g$ of types $f : A \to B$ and $g : B \to A$ such that $f(g\,x) \termeqrel x$ and $g(f\,x) \termeqrel x$.
The first type isomorphism is $\alwaystype{\kappa}{A} \typeiso A$ whenever $\kappa$ is not free in $A$.
The terms $g = \lambda x.\alwaysterm{\kappa}{x}$ of type $A \to \alwaystype{\kappa}{A}$ and $f = \lambda x .
\alwaysapp{x}{\kappa_0}$ of type $A \to \alwaystype{\kappa}{A}$ witness the isomorphism.
Note that we used the clock constant $\kappa_0$ in an essential way.
The equality $f(g\,x) \termeqrel x$ follows using only the $\beta$ rule for clock application.
The equality $g(f\,x) \termeqrel x$ follows using by the rule \typerule{TmEq-$\forall$-fresh}.

The following type isomorphisms follow by using $\beta$ and $\eta$ laws for the constructs involved.
\begin{itemize}
\item[-] If $\kappa \not \in A$ then $\alwaystype{\kappa}{\depprod{x}{A}{B}} \typeiso \depprod{x}{A}{\alwaystype{\kappa}{B}}$.
\item[-] $\alwaystype{\kappa}{\depsum{x}{A}{B}}
    \typeiso    \depsum{y}{\alwaystype{\kappa}{A}}{\left(\alwaystype{\kappa}{B\left[\alwaysapp{y}{\kappa}\right/x]}\right)}.$
\item[-] $\alwaystype{\kappa}{A} \typeiso \alwaystype{\kappa}{\later{\kappa}{A}}$.
\end{itemize}

There  is an important additional type isomorphism witnessing that $\forall\kappa$ commutes with binary sums; however unlike the
isomorphisms above we require equality reflection to show that the two functions are inverse to each other up to definitional equality.
There is a canonical term of type $\alwaystype{\kappa}{A} + \alwaystype{\kappa}{B} \to \alwaystype{\kappa}{(A + B)}$ using just ordinary elimination of coproducts.
Using the fact that we encode binary coproducts using $\Sigma$-types and universes we can define a term $\complus$ of type $\alwaystype{\kappa}{(A + B)} \to \alwaystype{\kappa}{A} + \alwaystype{\kappa}{B}$ which is a
inverse to the canonical term.
In particular $\complus$ satisfies the following two equalities which will be used below.
\begin{align}
  \label{eq:com-plus-equalities}
  \begin{split}
    \complus \left(\alwaysterm{\kappa}{\inl t}\right) &\termeqrel \inl \alwaysterm{\kappa}{t}
  \end{split}
  \begin{split}
    \complus \left(\alwaysterm{\kappa}{\inr t}\right) &\termeqrel \inr
    \alwaysterm{\kappa}{t}.
  \end{split}
\end{align}

\section{Example programs with coinductive types}
\label{sec:example-progr-coinductive}

Let $A$ be a type with code $\code{A}$ in clock context $\Delta$ and $\kappa$ a fresh clock variable.
Let $\stream{A} = \alwaystype{\kappa}\gstream[\kappa]{A}$.
We can define head, tail and cons functions
\begin{align*}
  \begin{split}
    \hd &: \stream{A} \to A\\
    \tl &: \stream{A} \to \stream{A}\\
    \cons &: A \to \stream{A} \to \stream{A}
  \end{split}
  \qquad
  \begin{split}
    \hd &\defeq \lambda xs.\hdg[\clockconst]\left(\alwaysapp{xs}{\clockconst}\right)\\
    \tl &\defeq \lambda xs.\prev{\kappa}{\tlg[\kappa]{(\alwaysapp{xs}{\kappa})}}\\
    \cons &\defeq \lambda x.\lambda
          xs.\alwaysterm{\kappa}\consg[\kappa]x\left(\pure{\kappa}{\left(\alwaysapp{xs}{\kappa}\right)}\right).
  \end{split}
\end{align*}

With these we can define the \emph{acausal} `every other' function $\everyother[\kappa]$ that removes every second element of the input stream.
It is acausal because the second element of the output stream is the third element of the input.
Therefore to type the function we need to have the input stream always available, so clock quantification must be used.
The function $\everyother[\kappa]$ of type $\stream{A} \to \gstream[\kappa]{A}$ is defined as
\begin{align*}
  \everyother[\kappa] &\defeq
                       \fix{\kappa}{\phi}{\lambda \left(xs : \stream{A}\right) .
                      \consg[\kappa] (\hd{xs})
                      \left(\phi \app[\kappa] \pure{\kappa}\left((\tl{(\tl{xs})})\right)\right)}.
\end{align*}
The result is a \emph{guarded} stream, but we can easily strengthen it and define $\everyother$ of type
$\stream{A} \to \stream{A}$ as
$\everyother \defeq \lambda xs.\alwaysterm{\kappa}{\everyother[\kappa] xs}$.

We can also work with covectors (not just guarded covectors as in Sec.~\ref{sec:examples}).
This is a dependent coinductive type indexed by conatural numbers which is the type
$\conat = \alwaystype{\kappa}{\gconat[\kappa]}$.
It is easy to define $\cozero$ and $\cosucc$ as $\cozero \defeq \alwaysterm{\kappa}{\inl\unit}$ and 
$\cosucc \defeq \lambda
    n.\alwaysterm{\kappa}{\inr{\left(\pure{\kappa}{\left(\alwaysapp{n}{\kappa}\right)}\right)}}$.
Next, we can define a transport function $\comnat$ of type $\comnat : \conat \to 1 + \conat$ satisfying
\begin{align}
  \label{eq:com-nat-equalities}
  \begin{split}
    \comnat \cozero &\termeqrel \inl\unit
  \end{split}
  \qquad
  \begin{split}
    \comnat (\cosucc n) &\termeqrel \inr n.
  \end{split}
\end{align}
This function is used to define the type family of covectors as
$\covec{A}\,n \defeq \alwaystype{\kappa}{\covec[\kappa]{A}\,n}$
where $\covec[\kappa]{A} : \conat \to \U{\Delta,\kappa}$ is the term
\begin{align*}
  \fix{\kappa}{\phi}{\lambda \left(n : \conat\right).}
  &\caseof{\comnat n}\left\{
  \caseinl{\_}{\code{1}};
  \caseinr{n}{A \code{\times} \latercode{\kappa}\left(\phi \app[\kappa] \left(\pure{\kappa}{n}\right)\right)}\right\}.
\end{align*}

Using term equalities \eqref{eq:com-plus-equalities} and \eqref{eq:com-nat-equalities} we can derive the type isomorphisms
\begin{align}
  \label{eq:vector-type-equality}
  \begin{split}
    \covec{A}\,\cozero &\typeeqrel \alwaystype{\kappa}{1} \typeiso 1\\
    \covec{A}\,(\cosucc n) &\typeeqrel \alwaystype{\kappa} \left(A \times
      \later{\kappa}\left(\covec[\kappa]{A}\,n\right)\right) \typeiso A \times \covec{A}\,n
  \end{split}
\end{align}
which are the expected properties of the type of covectors.

A simple function we can define is the tail function
\begin{align*}
  \begin{split}
    \covectail &: \covec{A}(\cosucc n) \to \covec{A}
  \end{split}
  \begin{split}
    \covectail &\defeq \lambda
    v.\prev{\kappa}{\pi_2\left(\alwaysapp{v}{\kappa}\right)}.
  \end{split}
\end{align*}
Note that \eqref{eq:vector-type-equality} is needed to type $\covectail$.
The $\covecmap$ function of type 
\begin{align*}
  \covecmap &: (A \to B) \to \depprod{n}{\conat}{\covec{A}n \to \covec{B}n}
\end{align*}
is defined as $\covecmap f \defeq \lambda n.\lambda xs.\alwaysterm{\kappa}{\covecmap[\kappa]f\,n\,\left(\alwaysapp{xs}{\kappa}\right)}$ where $\covecmap[\kappa]$ is
\begin{align*}
  \covecmap[\kappa] &: (A \to B) \to \depprod{n}{\conat}{\covec[\kappa]{A}n \to \covec[\kappa]{B}n}\\
  \covecmap[\kappa] &= \lambda f.
  \fix{\kappa}{\phi}
  \lambda n.\caseof{\comnat n}\\
            &\hspace{2.45cm}\caseinl{\_}{\lambda v.v}\\
            &\hspace{2.45cm}\caseinr{n}{\lambda v.\pair{f(\pi_1v)}{\phi \app[\kappa] (\pure{\kappa}{n}) \app[\kappa] \pi_2(v)}}.
\end{align*}

\subsection{Lifting guarded functions}
\label{sec:lift-guard-funct}

In this section we show how in general we may lift a function on guarded recursive types, such as addition of guarded streams, to a function on coinductive streams.
Moreover, we show how to lift proofs of properties, such as the commutativity of addition, from guarded recursive types to coinductive types.

Let $\Gamma$ be a context in clock context $\Delta$ and $\kappa$ a fresh clock.
Suppose $A$ and $B$ are types such that $\wftype{\Delta,\kappa}{\Gamma}{A}$ and $\wftype{\Delta,\kappa}{\Gamma, x : A}{B}$.
Finally let $f$ be a function of type $\hastype{\Delta,\kappa}{\Gamma}{f}{\depprod{x}{A}{B}}$.
We define $\Lf(f)$ satisfying the typing judgement
$\hastype{\Delta}{\Gamma}{\Lf(f)}{\depprod{y}{\alwaystype{\kappa}{A}}{\alwaystype{\kappa}{\left(B\left[\alwaysapp{y}{\kappa}/x\right]\right)}}}$
as $\Lf(f) \defeq \lambda y . \alwaysterm{\kappa}{f\left(\alwaysapp{y}{\kappa}\right)}$.

Now assume that $f'$ is another term of type $\depprod{x}{A}{B}$ (in the same context) and that we have proved
$\hastype{\Delta,\kappa}{\Gamma}{p}{\depprod{x}{A}{\idty{B}{f\,x}{f'\,x}}}$.
As above we can give the term $\Lf(p)$ the type 
$\depprod{y}{\alwaystype{\kappa}{A}}
  {\alwaystype{\kappa}
  {\idty{B\left[\alwaysapp{y}{\kappa}/x\right]}
        {f(\alwaysapp{y}{\kappa})}
        {f'(\alwaysapp{y}{\kappa})}}}.$
which by using the type equality \typerule{TyEq-$\forall$-Id} and the $\eta$ rule for $\forall$ is equal to the type
$\depprod{y}{\alwaystype{\kappa}{A}}
  {\idty{\alwaystype{\kappa}{B\left[\alwaysapp{y}{\kappa}/x\right]}}
        {\Lf(f)\,y}
        {\Lf(f')\,y}}$.
So we have derived a property of lifted functions $\Lf(f)$ and $\Lf(f')$ from the properties of the guarded versions $f$ and $f'$.
This is a standard pattern.
Using L\"ob induction we prove a property of a function whose result is a ``guarded'' type and derive the property for the lifted function.

For example we can lift the $\zipWith{}$ function from guarded streams to coinductive streams and prove that it preserves commutativity, using
the result on guarded streams of Sec.~\ref{sec:zipwith-preserves-comm}.

\section{Soundness}
\label{sec:soundness}

$\gdtt$ can be shown to be sound with respect to a denotational model interpreting the type theory. 
The model is a refinement of Bizjak and M\o{}gelberg's~\cite{Bizjak-Mogelberg:clock-synchronisation} 
but for reasons of space we leave the description of a full model of $\gdtt$ for future work.
Instead, to provide some intuition for the semantics of delayed substitutions, we just describe how to interpret the rule
\begin{align}
  \label{eq:soundness:example-rule}
  \begin{split}
    \inferrule* {\wftype{}{x : A}{B} \\ \hastype{}{\empty}{t}{\later{}{A}}}
    {\wftype{}{\empty}{\later{}[\hrt{x\gets t}]B}}
  \end{split}
\end{align}
in the case where we only have one clock available.

The subsystem of $\gdtt$ with only one clock can be modelled in the category $\trees$, known as the topos of trees~\cite{Birkedal-et-al:topos-of-trees}, the presheaf category over the first infinite ordinal
$\omega$. The objects $X$ of $\trees$ are families of sets $X_1,X_2,\ldots$ indexed by the positive integers, together with families of \emph{restriction functions} $r_i^X: X_{i+1}\to X_i$ indexed similarly.
There is a functor $\latermod : \trees \to \trees$ which maps an object $X$ to the object
\[
\begin{smalldiagram}
  1 & \ar{l}[above]{!} X_1 & \ar{l}[above]{r_1^X} X_2 & \ar{l}[above]{r_2^X} X_3 & \ar{l}\cdots
\end{smalldiagram}
\]
where $!$ is the unique map into the terminal object.

In this model, a closed type $A$ is interpreted as an object of $\trees$ and the type $\wftype{}{x : A}{B}$ 
is interpreted as an indexed family of sets $B_i(a)$, for $a$ in $A_i$ together with maps 
$r_i^B(a)\co B_{i+1}(a) \to B_{i}(r_i^A(a))$. 
The term $t$ in \eqref{eq:soundness:example-rule} is interpreted as a morphism
$t : 1 \to \later{}{A}$ so $t_i(\ast)$ is an element of $A_i$ (here we write $\ast$ for the element of $1$).

The type $\wftype{}{\empty}{\later{}[\hrt{x\gets t}]B}$ is then interpreted as the object $X$, defined by 
\begin{align*}
  X_1  &= 1  & X_{i+1} &= B_i(t_{i+1}(\ast)). 
\end{align*}
Notice that the delayed substitution is interpreted by substitution (reindexing) in the model; the change of the index in the model ($B_{i}$ is reindexed along $t_{i+1}(\ast)$) corresponds to the delayed substitution in the type theory.
Further notice that if $B$ does not depend on $x$, then the interpretation of $\wftype{}{\empty}{\later{}[\hrt{x\gets t}]B}$ reduces to the interpretation $\later{}{B}$, which is defined to be $\latermod$ applied to the interpretation of $B$.

The above can be generalised to work for general contexts and sequences of delayed substitutions, and one can then validate that the definitional equality rules do indeed hold in this model.

\section{Related Work}
\label{sec:related}

Birkedal et al.~\cite{Birkedal-et-al:topos-of-trees} introduced dependent type theory with the $\laterbare$ modality, with semantics in the topos of trees.
The guardedness requirement was expressed using the syntactic check that every occurrence of a type variable lies beneath a $\laterbare$.
This requirement was subsequently refined by Birkedal and M{\o}gelberg~\cite{Birkedal:intensional}, who showed that guarded recursive types could be constructed via fixed-points of functions on universes.
However, the rules considered in these papers do not allow one to apply terms of type $\laterbare(\Pi(x : A) .
B)$, as the applicative functor construction $\app$ was defined only for simple function spaces.
They are therefore less expressive for both programming (consider the covector $\operatorname{ones}$, and function $\map$, of Sec.~\ref{sec:an-example-with-covectors}) and proving, noting the extensive use of delayed substitutions in our example proofs.
They further do not consider coinductive types, and so are restricted to causal functions.

The extension to coinductive types, and hence acausal functions, is due to Atkey and McBride~\cite{Atkey:Productive}, who introduced \emph{clock quantifiers} into a simply typed setting with guarded recursion.
M{\o}gelberg~\cite{Mogelberg:tt-productive-coprogramming} extended this work to dependent types and Bizjak and M{\o}gelberg~\cite{Bizjak-Mogelberg:clock-synchronisation} refined the model further to allow clock synchronisation.

Clouston et al.~\cite{Clouston:Programming} introduced the logic
$\logiclambdanext$ to prove properties of terms
of the (simply typed) guarded $\lambda$-calculus, $\lambdanext$. This allowed
proofs about coinductive types, but not in the integrated
fashion supported by dependent type theories. Moreover it
relied on types being ``total'', a property that in a dependently typed setting would entail a strong
elimination rule for $\laterbare$, which would lead
to inconsistency.

Sized types~\cite{Hughes:Proving} have been combined with copatterns~\cite{Abel:Wellfounded} as an alternative type-based approach for modular programming with coinductive types.
This work is more mature than ours with respect to implementation and the demonstration of syntactic properties such as normalisation, and so further development of $\gdtt$ is essential to enable proper comparison.
One advantage of $\gdtt$ is that the later modality is useful for examples beyond coinduction, and beyond the utility of sized types, such as the guarded recursive domain equations used to model program logics~\cite{Svendsen:Impredicative}.

\section{Conclusion and Future Work}
\label{sec:conclusion}

We have described the dependent type theory $\gdtt$.
The examples we have detailed show that $\gdtt$ provides a setting for programming and proving with guarded recursive and coinductive types.

In future work we plan to investigate an intensional version of the type theory and construct a prototype implementation to allow us to experiment with larger examples.
Preliminary work has suggested that the path type of cubical type theory~\cite{Cohen:Cubical} interacts better with the new constructs
of $\gdtt$ than the ordinary Martin-L\"of identity type.

Finally, we are investigating whether the generalisation of applicative functors~\cite{McBride:Applicative}
to apply over \emph{dependent} function spaces, via delayed substitutions, might also apply to examples
quite unconnected to the later modality.

\paragraph{Acknowledgements.}
This research was supported in part by the ModuRes Sapere Aude Advanced Grant and 
DFF-Research Project 1 Grant no. 4002-00442, 
both from The Danish Council for Independent Research for the Natural Sciences (FNU).
Ale\v{s} Bizjak was supported in part by a Microsoft Research PhD grant.

\bibliographystyle{splncs03}
\bibliography{main}

\clearpage

\appendix

\begin{center}
  \Huge Appendix
\end{center}

\section{Overview of the appendix}
\label{app:sec:overview}

Sec.~\ref{app:sec:typingrules} contains type and term equalities of Fig.~\ref{fig:type-term-equalities} in full detail.
Sec.~\ref{app:sec:examples} starting on page~\pageref{app:sec:examples} contains detailed explanations of examples from Sec.~\ref{sec:examples} explaining how the rules of $\gdtt$ are used.
Sec.~\ref{app:sec:example-progr-coinductive} starting on page~\pageref{app:sec:example-progr-coinductive} contains detailed explanations of examples with coinductive types.
Sec.~\ref{app:sec:type-isom-details} starting on page~\pageref{app:sec:type-isom-details} contains a detailed derivation of the type isomorphism $\forall\kappa.A + B \typeiso \forall\kappa.A + \forall\kappa.B$ used in Sec.~\ref{sec:coinductive-types}.

\section{Typing rules}
\label{app:sec:typingrules}
\addcontentsline{toc}{section}{Typing rules}

  \paragraph{Definitional type equalities:}
  \begin{mathpar}
    \inferrule*[right={TyEq-$\later{}{}$-Weak}]
    {
      \wftype{\Delta}{\Gamma,\Gamma'}{A} \\
      \dsubst{\Delta}{\kappa}{\xi[x\gets t]}{\Gamma}{\Gamma',x:B}
    }
    {%
      \typeeq{\Delta}{\Gamma}{\later{\kappa}[\xi\hrt{x\gets t}]A}{\later{\kappa}[\xi]{A}}
    }
    \and
    \inferrule*[Right=TyEq-$\later{}$-Exch]
    {
      \wftype{\Delta}{\Gamma,\Gamma',x:B,y:C,\Gamma''}{A} \\
      \dsubst{\Delta}{\kappa}{\xi\hrt{x\gets t,y\gets u}\xi'}{\Gamma}{\Gamma',x:B,y:C,\Gamma''} \\
      \text{$x$ not free in $C$}
    }
    {%
      \typeeq{\Delta}{\Gamma}
      {\later{\kappa}[\xi\hrt{x\gets t,y\gets u}\xi']{A}}
      {\later{\kappa}[\xi\hrt{y\gets u,x\gets t}\xi']{A}}
    }
    \and
    \inferrule*[right={TyEq-Force}]
    {
      \wftype{\Delta}{\Gamma}{\later{\kappa}[\xi\hrt{x\gets\pure{\kappa}[\xi]{t}}]{A}}
    }
    {%
      \typeeq{\Delta}{\Gamma}
      {\later{\kappa}[\xi\hrt{x\gets\pure{\kappa}[\xi]{t}}]{A}}
      {\later{\kappa}[\xi]{A\subst{t}{x}}}
    }
    \and
    \inferrule*[right={TyEq-El-$\later{}$}]
    {
      \Delta' \subseteq \Delta \\ \isclock{\Delta'}{\kappa} \\
      \hastype{\Delta}{\Gamma, \Gamma'}{A}{\U{\Delta'}} \\
      \dsubst{\Delta}{\kappa}{\xi}{\Gamma}{\Gamma'}
    }
    {%
      \typeeq{\Delta}{\Gamma}
      {\El(\latercode{\kappa}\left(\pure{\kappa}[\xi]{A}\right))}
      {\later{\kappa}[\xi]\El(t)}
    }
    \and
    \inferrule*[right={TyEq-$\later{}{}$}]
    {
      \dsubst{\Delta}{\kappa}{\xi}{\Gamma}{\Gamma'}\\
      \hastype{\Delta}{\Gamma, \Gamma'}{t}{A} \\
      \hastype{\Delta}{\Gamma, \Gamma'}{s}{A}
    }
    {%
      \typeeq{\Delta}{\Gamma}
      {\idty{\later{\kappa}[\xi]{A}}{\pure{\kappa}[\xi]{t}}{\pure{\kappa}[\xi]{s}}}
      {\later{\kappa}[\xi]{\idty{A}{t}{s}}}
    }
  \end{mathpar}

  \paragraph{Definitional term equalities:}
  \begin{mathpar}
    \inferrule*[right={TmEq-Next-Weak}]
    {
      \hastype{\Delta}{\Gamma, \Gamma'}{u}{A} \\
      \dsubst{\Delta}{\kappa}{\xi\hrt{x \gets t}}{\Gamma}{\Gamma', x : B}
    }
    {%
      \termeq{\Delta}{\Gamma}
      {\pure{\kappa}[\xi\hrt{x \gets t}]{u}}
      {\pure{\kappa}[\xi]{u}}
      {\later{\kappa}[\xi]{A}}
    }
    \and
    \inferrule*[right={TmEq-Next-Var}]
    {
      \hastype{\Delta}{\Gamma}{t}{\later{\kappa}[\xi]{A}}
    }
    {%
      \termeq{\Delta}{\Gamma}{\pure{\kappa}[\xi\hrt{x\gets t}]{x}}{t}{\later{\kappa}[\xi]{A}}
    }
    \and
    \inferrule*[right={TmEq-Next-Exch}]
    {
      \hastype{\Delta}{\Gamma,\Gamma',x:B,y:C,\Gamma''}{t}{A} \\
      \dsubst{\Delta}{\kappa}{\xi\hrt{x\gets t,y\gets u}\xi'}
      {\Gamma}{\Gamma',x:B,y:C,\Gamma''} \\
      \text{$x$ not free in $C$}
    }
    {%
      \termeq{\Delta}{\Gamma}
      {\pure{\kappa}[\xi\hrt{x \gets t, y \gets u}\xi']{v}}
      {\pure{\kappa}[\xi\hrt{y \gets u, x \gets t}\xi']{v}}
      {\later{\kappa}[\xi\hrt{y \gets u, x \gets t}\xi']{A}}
    }
    \and
    \inferrule*[right={TmEq-Force}]
    {
      \hastype{\Delta}{\Gamma}{\pure{\kappa}[\xi\hrt{x \gets \pure{\kappa}[\xi]{t}}]{u}}
      {\later{\kappa}[\xi\hrt{x \gets \pure{\kappa}[\xi]{t}}]{A}}
    }
    {%
      \termeq{\Delta}{\Gamma}
      {\pure{\kappa}[\xi\hrt{x \gets \pure{\kappa}[\xi]{t}}]{u}}
      {\pure{\kappa}[\xi]{u\subst{t}{x}}}
      {\later{\kappa}[\xi]{A\subst{t}{x}}}
    }
    \and
    \inferrule*[right={TmEq-Fix}]
    {
      \hastype{\Delta}{\Gamma}{\fix{\kappa}{x}{t}}{A}
    }
    {%
      \termeq{\Delta}{\Gamma}
      {\fix{\kappa}{x}{t}}
      {t\subst{\pure{\kappa}{\left(\fix{\kappa}{x}{t}\right)}}{x}}
      {A}
    }
  \end{mathpar}

\section{Examples}
\label{app:sec:examples}
In this section we provide detailed explanations of typing derivations of examples described in Sec.~\ref{sec:examples}.

\subsection{$\zipWith{\kappa}$ preserves commutativity}
\label{app:sec:zipwith-preserves-comm}

The first proof is the simplest.
We will define the standard $\zipWith{\kappa}$ ($\operatorname{zipWith}$) function on streams and show that if a binary function $f$ is commutative, then so is $\zipWith{\kappa} f$.

The $\zipWith{\kappa} : (A \to B \to C) \to \gstream[\kappa]{A} \to \gstream[\kappa]{B} \to \gstream[\kappa]{C}$ is defined by guarded recursion as
\begin{align*}
  \zipWith{\kappa} f \defeq \fix{\kappa}{\phi}{\lambda& (xs, ys :
                                                        \gstream[\kappa]{A}) .  \\
                    &\consg[\kappa] \left(f\, (\hdg[\kappa]{xs})\,(\hdg[\kappa]{ys})\right)
                    \left(\phi\app[\kappa] \tlg[\kappa]{xs} \app[\kappa] \tlg[\kappa]{ys}\right)}
\end{align*}
Note that none of the new generalised $\later{}{}$ rules of $\gdtt$ are needed to type this function; this is a function on simple types.

Where we need dependent types is, of course, to state and prove properties.
To prove our example, that commutativity of $f$ implies commutativity of 
$\zipWith{\kappa} f$, means we must show that the type
\begin{align*}
  &\Pi (f : A \to A \to B) . \left(\depprod{x,y}{A}{\idty{B}{f\,x\,y}{f\,y\,x}}\right) \to\\
  &\depprod{xs,ys}{\gstream[\kappa]{A}}
  {\idty{\gstream[\kappa]{B}}{\zipWith{\kappa} f\,xs\,ys}{\zipWith{\kappa} f\,ys\,xs}}.
\end{align*}
is inhabited. We will explain how to construct such a term, and why it is typeable in
$\gdtt$. Although this construction might appear complicated at first, the actual
proof term that we construct will be as simple as possible.

Let $f : A \to A \to B$ be a function and say we have a term
\begin{align*}
  c : \depprod{x,y}{A}{\idty{B}{f\,x\,y}{f\,y\,x}}
\end{align*}
witnessing commutativity of $f$. We now wish to construct a term of type
\begin{align*}
  \depprod{xs,ys}{\gstream[\kappa]{A}}
  {\idty{\gstream[\kappa]{C}}{\zipWith{\kappa} f\,xs\,ys}{\zipWith{\kappa} f\,ys\,xs}}
\end{align*}
We do this by guarded recursion. To this end we assume
\begin{align*}
  \phi : \later{\kappa}\left(\depprod{xs,ys}{\gstream[\kappa]{A}}
  {\idty{\gstream[\kappa]{B}}{\zipWith{\kappa} f\,xs\,ys}{\zipWith{\kappa} f\,ys\,xs}}\right)
\end{align*}
and take $xs, ys : \gstream[\kappa]{A}$. Using $c$ (the proof that $f$ is commutative) we first
have $c\,(\hdg[\kappa]{xs})\,(\hdg[\kappa]{ys})$ of type 
\begin{align*}
  \idty{B}{f\,(\hdg[\kappa]{xs})\,(\hdg[\kappa]{ys})}{f\,(\hdg[\kappa]{ys})\,(\hdg[\kappa]{xs})}
\end{align*}
and because we have by definition of $\zipWith{\kappa}$
\begin{align*}
  \hdg[\kappa]{(\zipWith{\kappa} f\,xs\,ys)} &\judgeq f\,(\hdg[\kappa]{xs})\,(\hdg[\kappa]{ys})\\
  \hdg[\kappa]{(\zipWith{\kappa} f\,ys\,xs)} &\judgeq f\,(\hdg[\kappa]{ys})\,(\hdg[\kappa]{xs})
\end{align*}
we see that $c\,(\hdg[\kappa]{xs})\,(\hdg[\kappa]{ys})$ has type
\begin{align*}
   \idty{B}{\hdg[\kappa]{(\zipWith{\kappa} f\,xs\,ys)}}{\hdg[\kappa]{(\zipWith{\kappa} f\,ys\,xs)}}.
\end{align*}
To show that the tails are equal we use the induction hypothesis $\phi$.
The terms $\tlg[\kappa]{xs}$ and $\tlg[\kappa]{ys}$ are of type $\later{\kappa}{\gstream[\kappa]{A}}$,
so we first have $\phi \app[\kappa] \tlg[\kappa]{xs}$ of type
\begin{align*}
  \later{\kappa}[\hrt{xs\gets\tlg[\kappa]{xs}}]{\left(
    \Pi \left(ys : \gstream[\kappa]{A}\right).
    \idtyvert{\gstream[\kappa]{C}}{\zipWith{\kappa} f\,xs\,ys}{\zipWith{\kappa} f\,ys\,xs}\right)}
\end{align*}
Note the appearance of the generalised $\later{}{}$, carrying a delayed substitution.
Because the variable $xs$ does not appear in $\later{\kappa}{\gstream[\kappa]{A}}$ we may apply the weakening rule \typerule{TmEq-Next-Weak} to derive
\begin{align*}
  \tlg[\kappa]{ys} : \later{\kappa}[\hrt{xs \gets \tlg[\kappa]{xs}}]\gstream[\kappa]{A}
\end{align*}
Hence we may use the derived applicative rule to have $\phi \app[\kappa] \tlg[\kappa]{xs} \app[\kappa] \tlg[\kappa]{ys}$ of type
\begin{align*}
  \later{\kappa}[\vrt{xs\gets\tlg[\kappa]{xs}\\ ys\gets\tlg[\kappa]{ys}}]
  {\idty{\gstream[\kappa]{C}}{\zipWith{\kappa} f\,xs\,ys}{\zipWith{\kappa} f\,ys\,xs}}
\end{align*}
and which is definitionally equal to the type
\begin{align*}
  \idtyvert{\later{\kappa}\gstream[\kappa]{C}}
  {\pure{\kappa}[\vrt{xs\gets\tlg[\kappa]{xs}\\ ys\gets\tlg[\kappa]{ys}}]\zipWith{\kappa} f\,xs\,ys}
  {\pure{\kappa}[\vrt{xs\gets\tlg[\kappa]{xs}\\ ys\gets\tlg[\kappa]{ys}}]\zipWith{\kappa} f\,ys\,xs}.
\end{align*}
We also compute
\begin{align*}
  \tlg[\kappa]{(\zipWith{\kappa} f\,xs\,ys)}
  &\judgeq \pure{\kappa}{(\zipWith{\kappa} f)} \app[\kappa] \tlg[\kappa]{xs} \app[\kappa] \tlg[\kappa]{ys}\\
  &\judgeq \pure{\kappa}[\vrt{xs\gets \tlg[\kappa]{xs}\\ys\gets{\tlg[\kappa]{ys}}}]{(\zipWith{\kappa} f\,xs\,ys)}
\end{align*}
and
\begin{align*}
  \tlg[\kappa]{(\zipWith{\kappa} f\,ys\,xs)}
  &\judgeq \pure{\kappa}[\vrt{ys\gets \tlg[\kappa]{ys}\\zs\gets{\tlg[\kappa]{xs}}}]{(\zipWith{\kappa} f\,ys\,xs)}.
\end{align*}
Using the exchange rule \typerule{TmEq-Next-Exch} we have the equality
\begin{align*}
  \pure{\kappa}[\vrt{ys\gets \tlg[\kappa]{ys}\\xs\gets{\tlg[\kappa]{xs}}}]{(\zipWith{\kappa} f\,xs\,ys)} \equiv
  {\pure{\kappa}[\vrt{xs\gets \tlg[\kappa]{xs}\\ys\gets{\tlg[\kappa]{ys}}}]{(\zipWith{\kappa} f\,xs\,ys)}}.
\end{align*}
Putting it all together we have shown that the term $\phi \app[\kappa] \tlg[\kappa]{xs}\app[\kappa] \tlg[\kappa]{ys}$ has
type
\begin{align*}
  \idty{\later{\kappa}{\gstream[\kappa]{B}}}{\tlg[\kappa]{(\zipWith{\kappa} f\,xs\,ys)}}{\tlg[\kappa]{(\zipWith{\kappa} f\,ys\,xs)}}
\end{align*}
which means that the term
\begin{align*}
  \fix{\kappa}{\phi}{\lambda \left(xs, ys : \gstream[\kappa]{A}\right) .
  \pairseta\,\left(c\,(\hdg[\kappa]{xs})\,(\hdg[\kappa]{ys})\right)
                      \,\left(\phi\app[\kappa] \tlg[\kappa]{xs}\app[\kappa] \tlg[\kappa]{ys}\right)}
\end{align*}
has type
$\depprod{xs,ys}{\gstream[\kappa]{A}}
  {\idty{\gstream[\kappa]{B}}{\zipWith{\kappa} f\,xs\,ys}{\zipWith{\kappa} f\,ys\,xs}}.$

Notice that the resulting proof term could not be simpler than it is. In particular, we do
not have to write delayed substitutions in terms, but only in the intermediate types.

\subsection{An example with covectors}
\label{app:sec:an-example-with-covectors}

The next example is more sophisticated, as it will involve programming and proving with a data type that, unlike streams, is dependently typed.
In particular, we will see that the generalised later, carrying a delayed substitution, is necessary to type even the most elementary programs.

\emph{Covectors} are to colists (potentially infinite lists) as vectors are to lists.
To define guarded covectors we first need guarded co-natural numbers.
This is the type satisfying
\begin{align*}
  \gconat[\kappa] \typeeqrel \Unit + \later{\kappa}\gconat[\kappa].
\end{align*}
where binary sums are encoded in the type theory in a standard way.
The definition in $\gdtt$ is $\gconat[\kappa] \defeq \El\left(\fix{\kappa}{\phi}{(\code{\Unit} \codeop{+} \latercode{\kappa}\phi)}\right)$.

Using $\gconat[\kappa]$ we define the type of covectors of type $A$, written $\gcovec[\kappa]{A}$, as a $\gconat[\kappa]$-indexed type satisfying
\begin{align*}
  \gcovec[\kappa]{A}(\inl \unit) &\typeeqrel \Unit\\
  \gcovec[\kappa]{A}(\inr (\pure{\kappa}{m})) &\typeeqrel A \times \later{\kappa}{(\gcovec[\kappa]{A} m)}
\end{align*}
In $\gdtt$ we first define $\code{\gcovec[\kappa]{A}}$
\begin{align*}
  \code{\gcovec[\kappa]{A}} \defeq
  \fix{\kappa}{\phi}{\lambda (n : \gconat[\kappa]).
  &\caseof{n}\\
  &\caseinl{u}{\code{\Unit}}\\
  &\caseinr{m}{A \codeop{\times} \latercode{\kappa}(\phi \app[\kappa] m)}}.
\end{align*}
and then $\gcovec[\kappa]{A} n \defeq \El(\code{\gcovec[\kappa]{A}}\,n)$.
In the examples we will not distinguish between $\gcovec[\kappa]{A}$ and $\code{\gcovec[\kappa]{A}}$.
In the above $\phi$ has type $\later{\kappa}{(\gconat[\kappa] \to \U{\Delta,\kappa})}$ and inside the branches, $u$ has type $\Unit$ and $m$ has type $\later{\kappa}{\gconat[\kappa]}$, which is evident from the definition of $\gconat[\kappa]$.
As an example of covectors, we define $\tm{ones}$ of type $\Pi (n : \gconat[\kappa]) .
\gcovec[\kappa]{\NAT} n$ which produces a covector of any length consisting only of ones:
\begin{align*}
  \tm{ones} \defeq 
  & \fix{\kappa}{\phi}{\lambda (n : \gconat[\kappa]) . \caseof{n}} \\
  & \quad\caseinl{u}{\inl{\unit}} \\
  & \quad\caseinr{m}{\pair{1}{\phi \app[\kappa] m}}.
\end{align*}
When checking the type of this program, we need the generalised later.
The type of the recursive call is $\later{\kappa}{(\Pi (n : \gconat[\kappa]) .
  \gcovec[\kappa]{\NAT} n)}$, the type of $m$ is $\later{\kappa}{\gconat[\kappa]}$, and therefore the type of the subterm $\phi \app[\kappa] m$ must be
\begin{align*}
  \later{\kappa}[\hrt{n \gets m}]{\Pi (n : \gconat[\kappa]) . \gcovec[\kappa]{\NAT} n}.x
\end{align*}

We now aim to define the function $\map$ on covectors and show that it preserves
composition. Given two types $A$ and $B$ the $\map$ function has type
\begin{align*}
  \map : (A \to B) \to \depprod{n}{\gconat[\kappa]}{\gcovec[\kappa]{A}{n} \to \gcovec[\kappa]{B}{n}}.
\end{align*}
and is defined by guarded recursion as
\begin{align*}
  \map f \defeq
  &\fix{\kappa}{\phi}{\lambda (n : \gconat[\kappa]) .\\
  &\caseof{n}\\
  &\caseinl{u}{\lambda (x : 1) . x}\\
  &\caseinr{m}{\begin{array}{c}
                 \lambda\left(p : A \times \later{\kappa}[\hrt{n\gets m}](\gcovec[\kappa]{A} n)\right) . \\
                 \pair{f\left(\pi_1 p\right)}{\phi \app[\kappa] m \app[\kappa] (\pi_2p)}
               \end{array}}}
\end{align*}
Let us see why the definition has the correct type. First, the types of subterms are
\begin{align*}
  \phi &: \later{\kappa}{(\depprod{n}{\gconat[\kappa]}{\gcovec[\kappa]{A}{n} \to \gcovec[\kappa]{B}{n}})}\\
  u &: \Unit\\
  m &: \later{\kappa}{\gconat[\kappa]}
\end{align*}
Let $C = \gcovec[\kappa]{A}{n} \to \gcovec[\kappa]{B}{n}$, and write $C(t)$ for $C[t/n]$.
By the definition of $\gcovec[\kappa]{A}$ and $\gcovec[\kappa]{B}$ we
have $C(\inl u) \typeeqrel \Unit \to \Unit$, and so $\lambda (x : \Unit) . x$ has type $C(\inl u)$.

By the definition of $\gcovec[\kappa]{A}$ we have
\begin{align*}
  \gcovec[\kappa]{A}(\inr m)
  &\typeeqrel
  A \times \El\left(\latercode{\kappa}(\pure{\kappa}{(\gcovec[\kappa]{A})} \app[\kappa] m)\right)\\
  &\typeeqrel
  A \times \later{\kappa}[\hrt{n \gets m}]\left(\gcovec[\kappa]{A} n\right)
\end{align*}
and analogously for $\gcovec[\kappa]{B}(\inr m)$. Hence the type $C(\inr m)$ is convertible to
\begin{align*}
  \left(A \times \later{\kappa}[\hrt{n \gets m}]\left(\gcovec[\kappa]{A} n\right)\right) \to
  \left(B \times \later{\kappa}[\hrt{n \gets m}]\left(\gcovec[\kappa]{B} n\right)\right).
\end{align*}
Further, using the derived applicative rule we have
\begin{align*}
  \phi\app[\kappa] m : \later{\kappa}[\hrt{n\gets m}]C(n)
\end{align*}
and because $\pi_2 p$ in the second branch has type
\begin{align*}
  \later{\kappa}[\hrt{n\gets m}](\gcovec[\kappa]{A} n)
\end{align*}
we may use the (simple) applicative rule again to get
\begin{align*}
  \phi\app[\kappa] m \app[\kappa] (\pi_2 p) : \later{\kappa}[\hrt{n \gets m}](\gcovec[\kappa]{B} n)
\end{align*}
which allows us to type
\begin{align*}
  \lambda\left(p : A \times \later{\kappa}[\hrt{n\gets m}](\gcovec[\kappa]{A} n)\right) .
  \pair{f\left(\pi_1 p\right)}{\phi \app[\kappa] m \app[\kappa] \pi_2(p)}
\end{align*}
with type $C(\inr m)$.
Notice that we have made essential use of the more general applicative rule to apply $\phi\app[\kappa] m$ to $\pi_2 p$.
Using the strong (dependent) elimination rule for binary sums we can type the whole case construct with type $C(n)$, which is what we need to give $\map$ the desired type.

Now we will show that $\map$ so defined satisfies a basic property, namely that it
preserves composition in the sense that the type (in the context where we have types $A$,
$B$ and $C$)
\begin{align}
  \label{eq:covector-map-prop}
  \begin{split}
    &\Pi(f : A \to B) (g : B \to C) (n : \gconat[\kappa]) (xs : \gcovec[\kappa]{A} n) .\\
    &\qquad\idty{\gcovec[\kappa]{C}n}{\map g\, n(\map f\,n\,xs)}{\map (g \comp f)\,n\,xs}
  \end{split}
\end{align}
is inhabited. The proof is, of course, by L\"ob induction.

First we record some definitional equalities which follow directly by unfolding the definitions
\begin{align*}
  \map f\,(\inl u)\, x &\judgeq x\\
  \map f\,(\inr m)\, xs
  &\judgeq \pair{f\left(\pi_1 xs\right)}{\pure{\kappa}{(\map f)} \app[\kappa] m \app[\kappa] \pi_2(xs)}\\
  \judgeq \langle & f(\pi_1 xs) ,
  \pure{\kappa}[\vrt{n\gets m\\ ys\gets \pi_2 xs}]{(\map f\,n\,ys)}
  \rangle
\end{align*}
and so iterating these two equalities we get
\begin{align*}
  \map g\,(\inl u)\, (\map f (\inl u)\,x) &\judgeq x\\
  \map g\,(\inr m)\, (\map f (\inr m)\,xs) &\judgeq \pair{g(f(\pi_1 xs))}{s}
\end{align*}
where $s$ is the term
\begin{align*}
  \pure{\kappa}[\vrt{n\gets m\\ zs\gets {\pure{\kappa}[\vrt{n\gets m\\ ys\gets \pi_2 xs}]{(\map f\,n\,ys)}}}]{(\map g\,n\,zs)}
\end{align*}
which is convertible, by the rule \typerule{TmEq-Force}, to the term
\begin{align*}
  \pure{\kappa}[\vrt{n\gets m\\ ys\gets \pi_2 xs}]{(\map g\,n\,(\map f\,n\,ys))}.
\end{align*}
Similarly we have
\begin{align*}
  \map (g\comp f)\,(\inl u)\, x &\judgeq x
\end{align*}
and $\map (g \comp f)\,(\inr m)\, xs$ convertible to
\begin{align*}
  \pair{g(f(\pi_1 xs))}{\pure{\kappa}[\vrt{n\gets m\\ ys\gets \pi_2 xs}]{(\map (g\comp f)\,n\,ys)}}.
\end{align*}
Now let us get back to proving property \eqref{eq:covector-map-prop}.
Take $f : A \to B$, $g : B\to C$ and assume
\begin{align*}
  \phi : \later{\kappa}{\Pi (n : \gconat[\kappa]) (xs : \gcovec[\kappa]{A} n)}.
         \idty{\gcovec[\kappa]{C}n}{\map g\, n(\map f\,n\,xs)}{\map (g \comp f)\,n\,xs}
\end{align*}
We take $n : \gconat[\kappa]$ and write 
\begin{align*}
P(n) &=
\Pi (xs : \gcovec[\kappa]{A} n).\idty{\gcovec[\kappa]{C}n}{\map g\, n(\map f\,n\,xs)}{\map (g \comp f)\,n\,xs}.
\end{align*}
Then similarly as in the definition of $\map$ and the definitional equalities for $\map$ above we compute
\begin{align*}
  P(\inl u) \typeeqrel \Pi (xs : 1) . \idty{1}{xs}{xs}
\end{align*}
and so we have $\lambda (xs : 1) .\refl{1}{xs}$ of type $P(\inl u)$.

The other branch (when $n = \inr m$) is of course a bit more complicated.
As before we have
\begin{align}
  \label{eq:covec-defeq}
  \gcovec[\kappa]{A}(\inr m) \typeeqrel A \times \later{\kappa}[\hrt{n \gets m}]{\gcovec[\kappa]{A}n}
\end{align}
So take $xs$ of type $\gcovec[\kappa]{A}(\inr m)$.
We need to construct a term of type
\begin{align*}
  \idty{\gcovec[\kappa]{C}n}{\map g\, n(\map f\,n\,xs)}{\map (g \comp f)\,n\,xs}.
\end{align*}
First we have
$\refl{C}{g(f(\pi_1xs))}$ of type $\idty{C}{g(f(\pi_1xs))}{g(f(\pi_1xs))}$.
Then because $m$ is of type $\later{\kappa}{\gconat[\kappa]}$ we can use the induction hypothesis $\phi$ to get $\phi\app[\kappa] m$ of type
\begin{align*}
  \later{\kappa}[\hrt{n\gets m}]{}\Pi (xs : \gcovec[\kappa]{A} n).
  \idty{\gcovec[\kappa]{C}n}{\map g\, n(\map f\,n\,xs)}{\map (g \comp f)\,n\,xs}.
\end{align*}
Using \eqref{eq:covec-defeq} we have $\pi_2xs$ of type $\later{\kappa}[\hrt{n \gets m}]{\gcovec[\kappa]{A}n}$ and so we can use the applicative rule again to give $\phi\app[\kappa] m\app[\kappa] \pi_2 xs$ the type
\begin{align*}
  \later{\kappa}[\vrt{n\gets m\\ xs \gets \pi_2 xs}]
  \idtyvert{\gcovec[\kappa]{C}n}{\map g\, n(\map f\,n\,xs)}{\map (g \comp f)\,n\,xs}
\end{align*}
which by the rule \textsf{TyEq-$\later{}{}$} is the same as
\begin{align*}
  \idtyvert{D}
  {\pure{\kappa}[\vrt{n\gets m\\ xs \gets \pi_2 xs}]\left(\map g\, n(\map f\,n\,xs)\right)}
  {\pure{\kappa}[\vrt{n\gets m\\ xs \gets \pi_2 xs}]\left(\map (g \comp f)\,n\,xs\right)}
\end{align*}
where $D$ is the type $\later{\kappa}[\hrt{n\gets m}]{\gcovec[\kappa]{C}n}$.
Thus we can give to the term
\begin{align*}
  \lambda (xs : \gcovec[\kappa]{A}(\inr m)) .
  \pairseta \left(\refl{C}{g(f(\pi_1xs))}\right)
  \left(\phi\app[\kappa] m\app[\kappa] \pi_2 xs\right)
\end{align*}
the type $P(\inr m)$.
Using the dependent elimination rule for binary sums we get the final proof of property~\eqref{eq:covector-map-prop} as the term
\begin{align*}
  &\lambda (f : A \to B) (g : B \to C) . \fix{\kappa}{\phi}{\lambda (n : \gconat[\kappa]).\\
  &\caseof{n}\\
  &\caseinl{u}{\lambda (xs : 1) . \refl{1}{xs}}\\
  &\caseinr{m}{
  \lambda (xs : \gcovec[\kappa]{A}(\inr m)) .
   \pairseta \left(\refl{C}{g(f(\pi_1xs))}\right)
    \left(\phi\app[\kappa] m\app[\kappa] \pi_2 xs\right)}}
\end{align*}
which is as simple as could be expected.

\subsection{Lifting predicates to streams}
\label{app:sec:lift-pred-stre}

Let $P : A \to \U{\Delta}$ be a predicate on type $A$ and $\kappa$ a clock variable not in $\Delta$.
We can define a lifting of this predicate to a predicate $\lift[\kappa]{P}$ on streams of elements of type $A$.
The idea is that $\lift[\kappa]{P}xs$ will hold precisely when $P$ holds for all elements of the stream.
However we do not have access to all the element of the stream at the same time.
As such we will have $\lift[\kappa]{P}xs$ if $P$ holds for the first element of the stream $xs$ now, and $P$ holds for the second element of the stream $xs$ one time step later, and so on.
The precise definition uses guarded recursion:
\begin{align*}
  \lift[\kappa]{P} &: \gstream[\kappa]{A} \to \U{\Delta, \kappa}\\
  \lift[\kappa]{P} &\defeq \fix{\kappa}{\phi}{\lambda \left(xs : \gstream[\kappa]{A}\right) .
                     P\left(\hdg[\kappa]{xs}\right) \codeop{\times} \latercode{\kappa}\left(\phi \app[\kappa] \tlg[\kappa]{xs}\right)}.
\end{align*}
In the above term the subterm $\phi$ has type $\later{\kappa}{\left(\gstream[\kappa]{A} \to \U{\Delta,\kappa}\right)}$
and so because $\tlg[\kappa]{xs}$ has type $\later{\kappa}\gstream[\kappa]{A}$ we may form $\phi \app[\kappa] \tlg[\kappa]{xs}$ of
type $\later{\kappa}\U{\Delta,\kappa}$ and so finally $\latercode{\kappa}(\phi \app[\kappa] \tlg[\kappa]{xs})$ has type $\U{\Delta,\kappa}$
as needed.

To see that this makes sense, we have for a stream $xs : \gstream[\kappa]{A}$
\begin{align*}
  \El\left(\lift[\kappa]{P}\,xs\right) \typeeqrel
  \El\left(P\left(\hdg[\kappa]{xs}\right)\right) \times
  \El\left(\latercode{\kappa} \left(\pure{\kappa}{\lift[\kappa]{P}} \app[\kappa] \tlg[\kappa]{xs}\right)\right).
\end{align*}
Using delayed substitution rules we have
\begin{align*}
  \pure{\kappa}{\lift[\kappa]{P}} \app[\kappa] \tlg[\kappa]{xs} \judgeq
  \pure{\kappa}[\hrt{xs \gets \tlg[\kappa]{xs}}]{\left(\lift[\kappa]{P}\,xs\right)}
\end{align*}
which gives rise to the type equality
\begin{align*}
  \El(\latercode{\kappa}\pure{\kappa}{\lift[\kappa]{P}} \app[\kappa] \tlg[\kappa]{xs})
  \typeeqrel
  \El\left(\latercode{\kappa}\pure{\kappa}[\hrt{xs \gets \tlg[\kappa]{xs}}]{\left(\lift[\kappa]{P}\,xs\right)}\right).
\end{align*}
Finally, the type equality rule \textsc{TyEq-El-$\later{}{}$} gives us
\begin{align*}
  \El\left(\latercode{\kappa}\pure{\kappa}[\hrt{xs \gets \tlg[\kappa]{xs}}]\left(\lift[\kappa]{P}\,xs\right)\right)
  \typeeqrel\ 
  \later{\kappa}[\hrt{xs \gets \tlg[\kappa]{xs}}]\El(\lift[\kappa]{P}\,xs).
\end{align*}
All of these together then give us the type equality
\begin{align*}
  \El\left(\lift[\kappa]{P}\,xs\right) \typeeqrel
  \El(P\,(\hdg[\kappa]{xs})) \times \later{\kappa}[\hrt{xs \gets \tlg[\kappa]{xs}}]\El(\lift[\kappa]{P}\,xs).
\end{align*}
And so if $xs = \consg[\kappa] x\, (\pure{\kappa} ys)$ we can further simplify, using
rule \textsc{TyEq-Force}, to get
\begin{align*}
  \later{\kappa}[\hrt{xs \gets \pure{\kappa} ys}]\El(\lift[\kappa]{P}\,xs)
  \typeeqrel
    \later{\kappa}\left(\El(\lift[\kappa]{P}\,xs)[ys/xs]\right)
  \typeeqrel
  \later{\kappa}\El(\lift[\kappa]{P}\,ys)
\end{align*}
which then gives 
$\El(\lift[\kappa]{P} xs) \typeeqrel \El\left(P\, x\right) \times \later{\kappa}\El\left(\lift[\kappa]{P}\,ys\right)$
which is in accordance with the motivation given above.

Because $\lift[\kappa]{P}$ is defined by guarded recursion, we prove its properties by L\"ob
induction. In particular, we may prove that if $P$ holds on $A$ then $\lift[\kappa]{P}$ holds
on $\gstream[\kappa]{A}$, i.e., that the type
\begin{align*}
  \left(\depprod{x}{A}{\El\left(P\,x\right)}\right) \to
  \left(\depprod{xs}{\gstream[\kappa]{A}}{\El\left(\lift[\kappa]{P}\,xs\right)}\right)
\end{align*}
is inhabited (in a context where we have a type $A$ and a predicate $P$). Take
$p : \depprod{x}{A}{\El\left(P\,x\right)}$, and since we are proving by L\"ob induction we
assume the induction hypothesis later
\begin{align*}
  \phi : \later{\kappa}\left(\depprod{xs}{\gstream[\kappa]{A}}{\El\left(\lift[\kappa]{P}\,xs\right)}\right).
\end{align*}
Let $xs : \gstream[\kappa]{A}$ be a stream. By definition of $\lift[\kappa]{P}$ we have the type equality
\begin{align*}
  \El(\lift[\kappa]{P} xs) \typeeqrel \El\left(P\hdg[\kappa]{xs}\right)
  \times \later{\kappa}[\hrt{xs \gets \tlg[\kappa]{xs}}]{\El\left(\lift[\kappa]{P}\,xs\right)}
\end{align*}
Applying $p$ to $\hdg[\kappa]{xs}$ gives us the first component
\begin{align*}
  p(\hdg[\kappa]{xs}) : \El\left(P\left(\hdg[\kappa]{xs}\right)\right)
\end{align*}
and applying the induction hypothesis $\phi$ we have
\begin{align*}
  \phi\app[\kappa]\tlg[\kappa]{xs} : \later{\kappa}[\hrt{xs \gets \tlg[\kappa]{xs}}]\El(\lift[\kappa]{P}\,xs)
\end{align*}
Thus combining this with the previous term we have the proof of the lifting property as the term
\begin{align*}
  \lambda &\left(p : \depprod{x}{A}{\El\left(P\,x\right)}\right).\\
  &\fix{\kappa}{\phi}{\lambda \left(xs : \gstream[\kappa]{A}\right)
    \left\langle p\left(\hdg[\kappa]{xs}\right), \phi \app[\kappa] \tlg[\kappa]{xs}\right\rangle}.
\end{align*}

\section{Example programs with coinductive types}
\label{app:sec:example-progr-coinductive}

Let $A$ be some small type in clock context $\Delta$ and $\kappa$, a fresh clock variable.
Let $\stream{A} = \alwaystype{\kappa}\gstream[\kappa]{A}$.
We can define head, tail and cons functions
\begin{align*}
  \begin{split}
    \hd &: \stream{A} \to A\\
    \hd &\defeq \lambda xs.\hdg[\clockconst]\left(\alwaysapp{xs}{\clockconst}\right)
  \end{split}
  \qquad
  \begin{split}
    \tl &: \stream{A} \to \stream{A}\\
    \tl &\defeq \lambda xs.\prev{\kappa}{\tlg[\kappa]{(\alwaysapp{xs}{\kappa})}}
  \end{split}
\end{align*}
\begin{align*}
  \cons &: A \to \stream{A} \to \stream{A}\\
  \cons &\defeq \lambda x.\lambda
          xs.\alwaysterm{\kappa}\consg[\kappa]x\left(\pure{\kappa}{\left(\alwaysapp{xs}{\kappa}\right)}\right).
\end{align*}

With these we can define the \emph{acausal} `every other' function $\everyother[\kappa]$ that removes every second element of the input stream.
This is acausal because the second element of the output stream is the third element of the input.
Therefore to type the function we need to have the input stream always available, necessitating the use clock quantification.
The function $\everyother[\kappa]$ is
\begin{align*}
  \everyother[\kappa] &: \stream{A} \to \gstream[\kappa]{A}\\
  \everyother[\kappa] &\defeq
                       \fix{\kappa}{\phi}{\lambda \left(xs : \stream{A}\right) .\\
                      &\qquad\qquad\qquad\consg[\kappa] (\hd{xs})
                      \left(\phi \app[\kappa] \pure{\kappa}\left((\tl{(\tl{xs})})\right)\right)}.
\end{align*}
i.e., we return the head immediately and then recursively call the function on the stream with the first two elements removed.
Note that the result is a \emph{guarded} stream, but we can easily strengthen it and define $\everyother$ of type
$\stream{A} \to \stream{A}$ as
$\everyother \defeq \lambda xs.\alwaysterm{\kappa}{\everyother[\kappa] xs}$.

A more interesting type is the type of covectors, which is a refinement of the guarded type of covectors defined in Sec.~\ref{sec:an-example-with-covectors}.
First we define the type of co-natural numbers $\conat$ as
\begin{align*}
  \conat = \alwaystype{\kappa}{\gconat[\kappa]}.
\end{align*}
It is easy to define $\cozero$ and $\cosucc$ as
\begin{align*}
  \begin{split}
    \cozero &: \conat\\
    \cozero &\defeq \alwaysterm{\kappa}{\inl\unit}
  \end{split}
  \begin{split}
    \cosucc &: \conat \to \conat\\
    \cosucc &\defeq \lambda
    n.\alwaysterm{\kappa}{\inr{\left(\pure{\kappa}{\left(\alwaysapp{n}{\kappa}\right)}\right)}}
  \end{split}.
\end{align*}
Next, we will use type isomorphisms to define a transport function $\comnat$ of type $\comnat : \conat \to 1 + \conat$ as
\begin{align*}
  \comnat \defeq \lambda n.&\caseof{\complus n}\\
                           &\caseinl{u}{\inl \alwaysapp{u}{\kappa_0}}\\
                           &\caseinr{n}{\inr \prev{\kappa}{\alwaysapp{n}{\kappa}}}
\end{align*}
This function satisfies term equalities
\begin{align}
  \label{app:eq:com-nat-equalities}
  \begin{split}
    \comnat \cozero &\termeqrel \inl\unit
  \end{split}
  \qquad
  \begin{split}
    \comnat (\cosucc n) &\termeqrel \inr n.
  \end{split}
\end{align}

Using this we can define type of covectors $\covec{A}$ as
\begin{align*}
  \covec{A}\,n \defeq \alwaystype{\kappa}{\covec[\kappa]{A}\,n}
\end{align*}
where $\covec[\kappa]{A} : \conat \to \U{\Delta,\kappa}$ is the term
\begin{align*}
  \fix{\kappa}{\phi}{\lambda \left(n : \conat\right).}
  &\caseof{\comnat n}\\
  &\caseinl{\_}{\code{1}}\\
  &\caseinr{n}{A \code{\times} \latercode{\kappa}\left(\phi \app[\kappa] \left(\pure{\kappa}{n}\right)\right)}.
\end{align*}
Notice the use of $\comnat$ to transport $n$ of type $\conat$ to a term of type $1 + \conat$ which we can case analyse.
To see that this type satisfies the correct type equalities we need some auxiliary term equalities which follow from the way we have defined the terms.

Using term equalities \eqref{eq:com-plus-equalities} and \eqref{eq:com-nat-equalities} we can derive the (almost) expected type equalities
\begin{align}
  \label{app:eq:vector-type-equality}
  \begin{split}
    \covec{A}\,\cozero &\typeeqrel \alwaystype{\kappa}{1}\\
    \covec{A}\,(\cosucc n) &\typeeqrel \alwaystype{\kappa} \left(A \times
      \later{\kappa}\left(\covec[\kappa]\,n\right)\right)
  \end{split}
\end{align}
and using the type isomorphisms we can extend these type equalities to type isomorphisms
\begin{align*}
  \covec{A}\,\cozero &\typeiso 1\\
  \covec{A}\,(\cosucc n) &\typeiso A \times \covec{A}\,n
\end{align*}
which are the expected type properties of the covector type.

A simple function we can define is the tail function
\begin{align*}
  \covectail &: \covec{A}(\cosucc n) \to \covec{A}\\
  \covectail &\defeq \lambda v.\prev{\kappa}{\pi_2\left(\alwaysapp{v}{\kappa}\right)}.
\end{align*}
Note that we have used \eqref{app:eq:vector-type-equality} to ensure that $\covectail$ is type correct.

Next, we define the $\covecmap$ function on covectors.
\begin{align*}
  \covecmap &: (A \to B) \to \depprod{n}{\conat}{\covec{A}n \to \covec{B}n}\\
  \covecmap f &= \lambda n.\lambda xs.\alwaysterm{\kappa}{\covecmap[\kappa]f\,n\,\left(\alwaysapp{xs}{\kappa}\right)}
\end{align*}
where $\covecmap[\kappa]$ is the function of type
\begin{align*}
  \covecmap[\kappa] : (A \to B) \to \depprod{n}{\conat}{\covec[\kappa]{A}n \to \covec[\kappa]{B}n}
\end{align*}
defined as
\begin{align*}
  \lambda f.
  \fix{\kappa}{\phi}
  \lambda n.&\caseof{\comnat n}\\
            &\caseinl{\_}{\lambda v.v}\\
            &\caseinr{n}{\lambda v.\pair{f(\pi_1v)}{\phi \app[\kappa] (\pure{\kappa}{n}) \app[\kappa] \pi_2(v)}}.
\end{align*}
Let us see that this has the correct type.
Let $D_A(x)$ (and analogously $D_B(x)$) be the type
\begin{align*}
  \begin{split}
    D_A(x)
  \end{split}
  \quad \defeq
  \begin{split}
    &\caseof{x}\\
    &\caseinl{\_}{\code{1}}\\
    &\caseinr{n}{A \code{\times} \latercode{\kappa}\left(\left(\pure{\kappa}{\covec[\kappa]{A}}\right) \app[\kappa] \left(\pure{\kappa}{n}\right)\right)}.
  \end{split}
\end{align*}
where $x$ is of type $1 + \conat$.
Using this abbreviation we can write the type of $\covecmap[\kappa]$ as
\begin{align*}
  (A \to B) \to \depprod{n}{\conat}{D_A(\comnat n) \to D_B(\comnat n)}.
\end{align*}
Using this it is straightforward to show, using the dependent elimination rule for sums, as we did in Sec.~\ref{sec:an-example-with-covectors}, that $\covecmap[\kappa]$ has the correct type.
Indeed we have $D_A(\inl z) \typeeqrel 1$ and $D_A(\inr n) \typeeqrel A \times \later{\kappa}\left(\covec{A}n\right)$.

\section{Type isomorphisms in detail}
\label{app:sec:type-isom-details}

\begin{itemize}
\item If $\kappa \not\in A$ then $\alwaystype{\kappa}{A} \typeiso A$.
  The terms are $\lambda x .
  x\left[\clockconst\right]$ and $\lambda x.\alwaysterm{\kappa}{x}$.
  The rule \typerule{TmEq-$\forall$-fresh} is crucially needed to show that they constitute a type isomorphism.
\item If $\kappa \not \in A$ then $\alwaystype{\kappa}{\depprod{x}{A}{B}} \typeiso \depprod{x}{A}{\alwaystype{\kappa}{B}}$. The terms are
  \begin{align*}
    \lambda z.\lambda x. \alwaysterm{\kappa}{\alwaysapp{z}{\kappa}\,x}
  \end{align*}
  of type
  $\alwaystype{\kappa}{\depprod{x}{A}{B}} \to \depprod{x}{A}{\alwaystype{\kappa}{B}}$
  and
  \begin{align*}
    \lambda z.\alwaysterm{\kappa}{\lambda x.\alwaysapp{(z\,x)}{\kappa}}
  \end{align*}
  of type 
  $\depprod{x}{A}{\alwaystype{\kappa}{B}} \to \alwaystype{\kappa}{\depprod{x}{A}{B}}$.
\item
  $\alwaystype{\kappa}{\depsum{x}{A}{B}}
    \typeiso
    \depsum{y}{\alwaystype{\kappa}{A}}{\left(\alwaystype{\kappa}{B\left[\alwaysapp{y}{\kappa}\right/x]}\right)}.$
  The terms are
  \begin{align*}
    \lambda z.\pair{\alwaysterm{\kappa}{\pi_1\left(\alwaysapp{z}{\kappa}\right)}}
    {\alwaysterm{\kappa}{\pi_2\left(\alwaysapp{z}{\kappa}\right)}}
  \end{align*}
  of type
  \begin{align*}
    \alwaystype{\kappa}{\depsum{x}{A}{B}}
    \to 
    \depsum{y}{\alwaystype{\kappa}{A}}{\left(\alwaystype{\kappa}{B\left[\alwaysapp{y}{\kappa}\right/x]}\right)}
  \end{align*}
  and
  \begin{align*}
    \lambda z.\alwaysterm{\kappa}{\pair{\alwaysapp{(\pi_1\,z)}{\kappa}}{\alwaysapp{(\pi_2\,z)}{\kappa}}}
  \end{align*}
  of the converse type.
\item $\alwaystype{\kappa}{A} \typeiso \alwaystype{\kappa}{\later{\kappa}{A}}$. The terms are
  \begin{align*}
    \lambda z.\alwaysterm{\kappa}{\pure{\kappa}{(\alwaysapp{z}{\kappa})}}
  \end{align*}
  of type
  $\alwaystype{\kappa}{A} \to \alwaystype{\kappa}{\later{\kappa}{A}}$
  and
  \begin{align*}
    \lambda z.\prev{\kappa}{\left(\alwaysapp{z}{\kappa}\right)}
  \end{align*}
  of the converse type.
  The $\beta$ and $\eta$ rules for $\prev{\kappa}{}$ ensure that this pair of functions constitutes an isomorphism.
\end{itemize}

Using these isomorphisms we can construct an additional type isomorphism witnessing that $\forall\kappa$ commutes with binary sums.
Recall that we encode binary coproducts using $\Sigma$-types and universes in the standard way.
Given two codes $\code{A}$ and $\code{B}$ in some universe $\U{\Delta}$ we define
\begin{align*}
  \code{A} \code{+} \code{B} &: \U{\Delta}\\
  \code{A} \code{+} \code{B} &\defeq \depsum{b}{\Bool}{\bif{b}{\code{A}}{\code{B}}}
\end{align*}
and we write $A + B$ for $\El\left(\code{A} \code{+} \code{B}\right)$.
Suppose that $\Delta' \subseteq \Delta$ and $\kappa$ is a clock variable not in $\Delta$.
Suppose that $\wfctx{\Delta}{\Gamma}$ and that we have two codes $\code{A}, \code{B}$ satisfying
\begin{mathpar}
  \hastype{\Delta,\kappa}{\Gamma}{\code{A}}{\U{\Delta',\kappa}}
  \and
  \hastype{\Delta,\kappa}{\Gamma}{\code{B}}{\U{\Delta',\kappa}}
\end{mathpar}

We start with an auxiliary function $\comif$.
Let $b$ be some term of type $\Bool$.
We then define
\begin{align*}
  \comif[b] &: \alwaystype{\kappa}{\El\left(\bif{b}{\code{A}}{\code{B}}\right)}\\
           &\qquad \to \El\left(\bif{b}
           {\alwayscode\alwaysterm{\kappa}{\code{A}}}
           {\alwayscode\alwaysterm{\kappa}{\code{B}}}\right)\\
  \comif[b] &\defeq \bif{b}{\lambda x.x}{\lambda x.x}
\end{align*}
which is typeable due to the strong elimination rule for $\Bool$.

We now define the function $\complus$
\begin{align*}
  \complus &: \alwaystype{\kappa}{(A + B)} \to \alwaystype{\kappa}{A} + \alwaystype{\kappa}{B}\\
  \complus &\defeq \lambda z.
             \pair{\pi_1\left(\alwaysapp{z}{\clockconst}\right)}
             {\comif[\pi_1\left(\alwaysapp{z}{\clockconst}\right)] \left(\alwaysterm{\kappa}\pi_2\left(\alwaysapp{z}{\kappa}\right)\right)}.
\end{align*}
We need to check that the types are well-formed and the function well-typed.
The side condition $\wfctx{\Delta}{\Gamma}$ ensures that the types are well-formed.
To see that the function $\complus$ is well-typed we consider the types of subterms.
\begin{itemize}
\item[-] The term $z$ has type $\alwaystype{\kappa}{(A+B)}$.
\item[-] The term $\pi_1\left(\alwaysapp{z}{\clockconst}\right)$ has type $\Bool$.
\item[-] The term $\alwaysterm{\kappa}{\pi_2\left(\alwaysapp{z}{\kappa}\right)}$ has type
  \begin{align*}    \alwaystype{\kappa}
    {\El\left(\bif{\pi_1\left(\alwaysapp{z}{\kappa}\right)}{\code{A}}{\code{B}}\right)}
  \end{align*}
\item[-] From \typerule{TmEq-$\forall$-fresh} we get $\pi_1(\alwaysapp{z}{\clockconst}) \termeqrel \pi_1(\alwaysapp{z}{\kappa})$.
  Indeed, the term
  \begin{align*}
    \alwaysterm{\kappa}{\pi_1(\alwaysapp{z}{\kappa})}
  \end{align*}
  has type $\Bool$, which does not contain $\kappa$, and the required equality follows from \typerule{TmEq-$\forall$-fresh} and the $\beta$ rule for clock quantification.   
\item[-] Thus the term $\alwaysterm{\kappa}{\pi_2\left(\alwaysapp{z}{\kappa}\right)}$ has type
  \begin{align*}
    \alwaystype{\kappa}
    {\El\left(\bif{\pi_1\left(\alwaysapp{z}{\clockconst}\right)}{\code{A}}{\code{B}}\right)}
  \end{align*}
\item[-] And so the term
  \begin{align*}
    \comif[\pi_1\left(\alwaysapp{z}{\clockconst}\right)]
    \alwaysterm{\kappa}{\pi_2\left(\alwaysapp{z}{\kappa}\right)}
  \end{align*}
  has type
  \begin{align*}
    \El\left(\bif{\pi_1\left(\alwaysapp{z}{\clockconst}\right)}
    {\alwayscode\alwaysterm{\kappa}{\code{A}}}{\alwayscode\alwaysterm{\kappa}{\code{B}}}\right)
  \end{align*}
  which is exactly the type needed to typecheck the whole term.
\end{itemize}

For the term $\complus$ we can derive the following definitional term equalities.
\begin{align}
  \label{app:eq:com-plus-equalities}
  \begin{split}
    \complus \left(\alwaysterm{\kappa}{\inl t}\right) &\termeqrel \inl \alwaysterm{\kappa}{t}\\
    \complus \left(\alwaysterm{\kappa}{\inr t}\right) &\termeqrel \inr
    \alwaysterm{\kappa}{t}
  \end{split}
\end{align}

There is also a canonical term of type
\begin{align*}
  \alwaystype{\kappa}{A} + \alwaystype{\kappa}{B} \to \alwaystype{\kappa}{(A + B)}
\end{align*}
defined as
\begin{align*}
  \lambda z.\alwaysterm{\kappa}{ }
  &\caseof{z}\\
  &\caseinl{a}{\inl{(\alwaysapp{a}{\kappa})}}\\
  &\caseinl{b}{\inl{(\alwaysapp{b}{\kappa})}}.
\end{align*}
This term is inverse to $\complus$, although we require equality reflection to show that the two functions are inverses to each other.
Without equality reflection we can only prove they are inverses up to propositional equality.
The isomorphisms defined previously do not require equality reflection.

\end{document}